%% file: main.tex
\pdfoutput=1

\documentclass[conference]{IEEEtran}

\ifCLASSINFOpdf
\else
\fi

\usepackage{xspace}
\usepackage{ragged2e}
\usepackage{xurl}
\usepackage[table]{xcolor}
\usepackage{listings}
\usepackage{tikz}
\usepackage{graphicx}
\usetikzlibrary{calc,arrows.meta,positioning}
\usepackage{multirow}
\usepackage{amsmath}
\usepackage{tabularx}
\usepackage{subcaption}
\usepackage[most]{tcolorbox}
\usepackage{fontawesome5}
\usepackage{pifont}

\usepackage{makecell}

\newcolumntype{Y}{>{\raggedright\arraybackslash}X}

\newcommand{\tightbox}[2]{%
  {\setlength{\fboxsep}{1pt}\colorbox{#1}{#2}}%
}
\tikzset{
  flowArrow/.style={dashed, very thick, -{Stealth[length=2.2mm]}},
  dataArrow/.style={dashed, thick, -{Stealth[length=2.0mm]}},
  bubble/.style={
        circle,          %
        draw=black,      %
        fill=white,      %
        text=black,      %
        font=\scriptsize\bfseries, %
        inner sep=1pt,   %
        align=center     %
    }
}
\lstdefinestyle{javastyle}{
  language=Java,
  basicstyle=\ttfamily\scriptsize,
  columns=fullflexible,
  keepspaces=true,
  numbers=left,
  numberstyle=\tiny,
  numbersep=6pt,
  frame=single,
  framerule=0.4pt,
  breaklines=true,
  showstringspaces=false,
  tabsize=2,
  captionpos=b
}

\usepackage{algorithm}
\usepackage{algpseudocode}
\makeatletter
\renewcommand{\ALG@name}{Algorithm}
\algrenewcommand\ALG@beginalgorithmic{\small}
\makeatother
\AtBeginDocument{
  \algrenewcommand\algorithmicrequire{\textbf{Input:}}
  \algrenewcommand\algorithmicensure{\textbf{Output:}}
  \algrenewcommand\algorithmiccomment[1]{\hfill$\triangleright$ #1}
}

\usepackage{enumitem}
\usepackage{hhline}
\usepackage{colortbl}
\usepackage{booktabs}
\usepackage{placeins}

\definecolor{slotTarget}{RGB}{200,100,40}
\definecolor{codegreen}{rgb}{0,0.6,0}
\definecolor{codegray}{rgb}{0.5,0.5,0.5}
\definecolor{codepurple}{rgb}{0.58,0,0.82}
\definecolor{backcolour}{rgb}{0.97,0.97,0.97}

\definecolor{slotThreat}{rgb}{0.86,0.20,0.25}   %
\definecolor{slotExample}{rgb}{0.20,0.55,0.30}  %
\definecolor{slotCtx}{rgb}{0.90,0.55,0.10}      %
\definecolor{slotCallee}{rgb}{0.25,0.40,0.85}   %
\definecolor{slotBody}{rgb}{0.55,0.25,0.75}     %
\definecolor{slotSchema}{rgb}{0.30,0.60,0.70}   %
\definecolor{slotCond}{rgb}{0.20,0.45,0.80}     %
\definecolor{slotSlice}{rgb}{0.10,0.55,0.55}    %
\definecolor{slotCache}{rgb}{0.45,0.45,0.45}    %
\definecolor{slotSink}{rgb}{0.75,0.30,0.60}     %
\definecolor{slotManifest}{rgb}{0.90,0.55,0.10} %
\definecolor{slotDyn}{rgb}{0.20,0.45,0.80}      %
\definecolor{slotRule}{rgb}{0.20,0.55,0.30}     %

\lstdefinelanguage{json}{
    basicstyle=\small\ttfamily,
    numbers=none,
    numberstyle=\tiny,
    stepnumber=1,
    numbersep=8pt,
    showstringspaces=false,
    breaklines=true,
    frame=lines,
    string=[s]{"}{"},
    stringstyle=\color{blue},
    comment=[l]{:},
    commentstyle=\color{black},
}

\newcommand{\lvlCritical}{\CIRCLE\CIRCLE} 
\newcommand{\lvlHigh}{\CIRCLE\LEFTcircle}
\newcommand{\lvlModerate}{\CIRCLE\Circle}
\newcommand{\lvlLow}{\LEFTcircle\Circle}
\newcommand{\lvlIgnore}{\Circle\Circle}

\newcommand{\lvlbox}{\path[use as bounding box] (0.65ex,-0.5ex) rectangle (4.35ex,0.5ex);}

\newcommand{\lvl}[1]{%
  \tikz[baseline={([yshift=-0.55ex]current bounding box.center)}]{%
    \lvlbox
    \foreach \i in {1,2,3}{%
      \pgfmathparse{\i<=#1 ? "black" : "none"}%
      \edef\fillcol{\pgfmathresult}%
      \draw[line width=0.4pt, fill=\fillcol] (\i*1.25ex,0) circle (0.42ex);%
    }%
  }%
}

\newcommand{\lvlhead}{%
  \tikz[baseline={([yshift=-0.55ex]current bounding box.center)}]{%
    \lvlbox
    \foreach \i/\t in {1/E,2/T,3/S}{%
      \node[inner sep=0pt, font=\bfseries\itshape\scriptsize] at (\i*1.25ex,0) {\t};%
    }%
  }%
}

\newcommand{\sysname}{\textsc{Callix}\xspace}

\newcommand{\ignore}[1]{}
\newcolumntype{L}[1]{>{\RaggedRight\arraybackslash}p{#1}}
\newcommand{\group}[1]{\textbf{#1}}

\providecommand{\Letter}{\ding{41}}

\usepackage{wasysym}
\usepackage{hyperref}
\hypersetup{hidelinks}

\begin{document}
\title{Beyond Sensitive APIs: Detecting Access Control Flaws at Sensitive Callsites in Android Firmware}

\author{
\IEEEauthorblockN{
Haotian Deng,
Meng Hao,
Hanxiao Chen,
Xinyuan Qian,
Guowen Xu,
Hongwei Li\textsuperscript{\Letter},
Chaoshun Zuo\textsuperscript{\Letter}
}
\IEEEauthorblockA{
University of Electronic Science and Technology of China\\
\textit{h.deng@std.uestc.edu.cn, menghao303@gmail.com, xinyuanqian@outlook.com,} \\ \textit{\{hanxiao.chen, guowen.xu, hongweili, chaoshun.zuo\}@uestc.edu.cn}
}
}

\IEEEoverridecommandlockouts
\makeatletter\def\@IEEEpubidpullup{4\baselineskip}\makeatother
\IEEEpubid{\parbox{\columnwidth}{
		Network and Distributed System Security (NDSS) Symposium 2027\\
		22--26 March 2027, Seoul, Republic of Korea\\
		ISBN 978-1-970672-09-1\\  
		https://dx.doi.org/10.14722/ndss.2027.231311\\
		www.ndss-symposium.org
}
\hspace{\columnsep}\makebox[\columnwidth]{}}

\maketitle
\IEEEpeerreviewmaketitle

\begin{NoHyper}
  \renewcommand{\thefootnote}{\Letter}
  \footnotetext{Corresponding authors.}
\end{NoHyper}

\input{abstract}

\input{introduction}

\input{background_motivation}

\input{overview}

\input{detailed_design}

\input{evaluation}

\input{discussion}

\input{related_work}

\input{conclusion}

\input{ethics}

\section*{Acknowledgment}
We express our deepest gratitude to the reviewers for their constructive comments.
This work is supported by the National Natural Science Foundation of China under Grants U25B2079 and 62502079, the China Postdoctoral Science Foundation under Grant BX20240053, and the Sichuan Natural Science Foundation under Grant 2026NSFSC1463.

\bibliographystyle{IEEEtran}

\bibliography{references}

\input{appendix/appendix-prompts}

\end{document}

%% file: abstract.tex
\begin{abstract}
Existing studies of vulnerable pre-installed Android apps identify dangerous behavior by checking whether an app invokes an API from a predefined set of security-sensitive APIs.
These sets usually exclude generic APIs to avoid false positives; unfortunately, this misses a class of vulnerabilities in which a generic API, such as a database read, appears benign in general but becomes security-sensitive when operating on protected data, yet is left without sufficient access control.
While this class of vulnerabilities appears addressable via access control analysis, existing studies focus only on the presence of checks, not on whether those checks enforce sufficient protection for the operation at a given callsite.
In this work, we propose \sysname, which first assesses callsite sensitivity by jointly analyzing the invoked API functionality and the data involved in the operation, then recovers each security check on the execution path as an explicit security property, and finally introduces the notion of restriction context to decide whether these checks are sufficient for the identified sensitive operation.
We evaluated \sysname on $1{,}750$ distinct pre-installed apps from the $10$ top-tier Android vendors as of $2025$.
In total, \sysname discovered $819$ vulnerable paths across these vendors, $42.1\%$ of which allow reading or modifying private user data through generic database operations.
Manual validation of a random $60\%$ sample shows a precision of $95.1\%$.
Many of our findings have been confirmed and rewarded through vendor bug bounty programs.
\end{abstract}

%% file: introduction.tex
\section{Introduction}
\label{sec:intro}

Android vendors enrich their services with pre-installed apps for system utilities, hardware support, and vendor-specific services~\cite{wu2013impact,harvesting16}.
Unlike third-party apps from public markets (e.g., Google Play), these apps often hold elevated privileges and are embedded in the firmware, making them difficult or impossible for users to remove~\cite{gamba2020analysis,firmscope}.
However, the security consequences of a single vulnerability in pre-installed apps would be amplified across all smartphones shipped by the respective vendor.
While numerous approaches have been proposed and have uncovered a wide range of security flaws, including privilege escalation vulnerabilities~\cite{firmscope},
pre-installed apps remain a highly privileged and persistent attack surface in Android, exposing 
sensitive data or system capabilities to ordinary third-party apps~\cite{firmscope, patch22}.

Among the vulnerabilities continuously reported over the years, a case~\cite{news} disclosed by the Polish CERT in mid-$2025$ raises two limitations that remain largely overlooked in existing studies.
In particular, the vulnerable pre-installed app, shipped on popular Android phones, exposes a \texttt{ContentProvider} interface that allows any app on the device to extract the user's PIN code without dedicated permissions.
Unfortunately, this vulnerability is difficult to capture with existing analyses for two reasons.
First, the vulnerable operation is triggered through a generic provider API that is not inherently security-sensitive and is therefore often excluded from predefined sensitive API sets to avoid false positives~\cite{firmscope}.
However, such an API could become security-sensitive at a particular callsite when it operates on protected user data, i.e., a user's PIN code.
Second, although the vulnerable interface may contain access control logic, existing analyses either ignore such branch conditions~\cite{firmscope, zhao2020automatic} or treat them mainly as path constraints for reachability reasoning, such as in symbolic execution~\cite{yang2013appintent, arzt2015using, cadar2013symbolic}.
Such reasoning can only establish whether a path is feasible instead of whether the conditions along the path provide sufficient authorization for the sensitive behavior.\looseness=-1

Inspired by this newly reported vulnerability, we therefore aim to bridge the gap to detect such insufficient access controls on sensitive callsites in \texttt{ContentProvider} components, an essential interface for data sharing in Android.
To this end, we propose \sysname, a static analysis framework that aims to identify security-sensitive uses of generic system APIs from the operation they perform and the data involved, and determine whether the access controls along each path provide sufficient protection.
\sysname addresses this problem in three stages.
First, it identifies sensitive callsites by jointly characterizing the API's operation and the semantics of the data involved, rather than relying on API identity alone.
Second, it recovers each access control check from branch conditions as an explicit security property that captures both the intended validation target and the information used for validation.
Third, it introduces the notion of restriction context to determine whether recovered checks are sufficient for the identified operation and data.

We implemented a prototype of \sysname and evaluated it on $50$ firmware images from $10$ major Android vendors and $1{,}750$ distinct pre-installed apps.
\sysname{} identifies $819$ vulnerable paths across $424$ \texttt{ContentProvider}s, showing that the problem is not a rare corner case confined to isolated apps or vendors.
To assess the effectiveness and accuracy of these findings, we manually validate $60\%$, i.e., $492$, randomly sampled paths, where $468$ are confirmed true positives, yielding $95.1\%$ precision, and each confirmed path is reproduced with a working proof of concept.
We further test whether these advancements stem merely from the underlying LLM by transferring the same static evidence, taxonomies, and sufficiency criteria to a commercial coding agent on a balanced $500$-path benchmark.
Even on the same \texttt{claude-opus-5} backend, the transferred formulation remains $6.6$ F1 points behind \sysname, indicating that the criteria alone are insufficient to reproduce \sysname's performance and that their structured application also matters.
Finally, affected vendors have independently confirmed $278$ disclosed reports across $98$ providers as of this writing, with some subsequently rewarded or remediated.
The remaining reports are under review.

To summarize, this work makes the following contributions:
\begin{itemize}[leftmargin=*]
    \item \textbf{A novel problem.} 
    We report a new problem: insufficient access control at sensitive callsites in pre-installed \texttt{ContentProvider} components. 
    In contrast to API-level analyses, which only target security-sensitive APIs, we show that a generic API could become sensitive depending on the operation it performs and the data involved at a particular callsite, which has been overlooked in the literature.
    \item \textbf{A new analysis.}
    We propose \sysname, a static analysis framework that identifies sensitive callsites from operation and data semantics, recovers access control checks as explicit security properties, and introduces the notion of restriction context to evaluate whether the recovered checks provide sufficient protection along each sensitive path.
    \item \textbf{Large-scale empirical findings.}
    We evaluate \sysname on $1{,}750$ distinct pre-installed apps extracted from $50$ firmware images spanning $10$ major Android vendors. 
    \sysname identifies $819$ vulnerable paths across $424$ \texttt{ContentProvider}s. 
    The findings illustrate that the \texttt{ContentProvider} interface exposes functionality far beyond conventional data sharing, reaching privileged system operations that could be exploited for varied malicious purposes.
    Many of our findings have been confirmed by the corresponding vendors and awarded multiple bug bounties.
\end{itemize}

%% file: background_motivation.tex
\section{Background}
\label{sec:background}

\subsection{Pre-installed Apps in Android}
\label{sec:bg-preapps}
An Android firmware image typically contains hundreds of pre-installed apps, e.g., from $200$ to $500$ in $2019$~\cite{firmscope}, ranging from the AOSP framework apps maintained by Google to OEM-specific components for connectivity, device management, telephony, and platform UI.
Unlike third-party apps that can be downloaded from public app stores (e.g., Google Play), 
pre-installed apps ship as part of the firmware and may hold permissions unavailable to ordinary apps, allowing them to access privileged system APIs and implement core system or vendor functionality~\cite{gamba2020analysis,firmscope}.
Since they are integrated into the firmware, users usually cannot uninstall them through ordinary app-management mechanisms, and a flaw in such an app can persist on devices until the vendor distributes an update.

\subsection{\texttt{ContentProvider} in Android}
\label{sec:bg-provider}
The \texttt{ContentProvider} is Android's primary component for mediating cross-app data sharing. 
It exposes app-managed data through a unified \texttt{URI}-based interface, through which other apps or components can access provider-managed resources in a structured manner.
In addition, instead of directly exposing files, databases, or internal storage paths, the \texttt{ContentProvider} abstracts the underlying storage implementation and provides the following standardized operations for external access~\cite{android-contentprovider-doc}:
\begin{itemize}[leftmargin=*]
    \item \texttt{ContentProvider.query()} returns records that match a given \texttt{URI} to the caller.
    \item \texttt{ContentProvider.insert()} adds new records and returns the \texttt{URI} of the created resource.
    \item \texttt{ContentProvider.update()} modifies existing records identified by a \texttt{URI}.
    \item \texttt{ContentProvider.delete()} removes matching records from the underlying storage.
    \item \texttt{ContentProvider.openFile()} returns a file descriptor for the resource identified by the \texttt{URI}.
    \item \texttt{ContentProvider.getType()} returns the MIME type of the data identified by the given \texttt{URI}.
    \item \texttt{ContentProvider.call()} exposes provider-defined operations outside the standard CRUD semantics~\cite{aospprovider}.
\end{itemize}
Although these entry points are standardized, their concrete semantics are determined by the provider's implementation, which maps \texttt{URI}s, method names, and arguments to database records, files, privileged operations, or other application-specific resources.
Client apps typically interact with a \texttt{ContentProvider} through the Android \texttt{ContentResolver}.
Each \texttt{ContentProvider} is identified by a unique \texttt{authority} and exposes resources using \texttt{URI}s of the form \texttt{content://\allowbreak authority/\allowbreak path/\allowbreak id}.
Such \texttt{URI}s specify both the target provider (by \texttt{authority}) and the particular data resource (by \texttt{path}) to be accessed.
The Android framework is responsible for resolving the provider, dispatching the request through inter-process communication, and enforcing access control checks~\cite{android-contentprovider-doc, android-system-permission-doc}.

\subsection{Access Control for \texttt{ContentProvider}}
\label{sec:bg-protection}

In Android, \texttt{ContentProvider} can enforce its preferred access control statically via declaration in the \texttt{Manifest} file or dynamically via code implementation~\cite{android-system-permission-doc, provider}.

\noindent
\textbf{Static access control.}
There are several access restrictions that could be declared in the \texttt{Manifest}.
Specifically, developers can configure the attribute \texttt{android:exported} to restrict the provider's external accessibility, the attribute \texttt{android:permission} for identity authentication, and the attributes \texttt{android:readPermission} and \texttt{android:writePermission} to regulate read and write operations, respectively~\cite{provider}.
Developers can also use \texttt{<path-permission>} to assign permissions to specific \texttt{URI} paths for fine-grained access restrictions~\cite{provider}.
In addition, there are different permission levels.
In particular, \texttt{normal} permissions are granted automatically to any app that requests them at install time; \texttt{dangerous} permissions are runtime permissions granted only with explicit user consent; and \texttt{signature} permissions restrict access to apps signed with the same certificate as the app that declares the permission~\cite{android-system-permission-doc}.
To refine the conditions under which a permission may be granted, Android further supports optional protection-level flags such as \texttt{privileged}, \texttt{preinstalled}, and \texttt{development}.\looseness=-1

\noindent
\textbf{Dynamic access control.}
Unlike static access control, which is explicitly specified by Android documentation~\cite{provider}, such dynamic checks are largely implementation-defined and may vary across providers.
For example, it can check whether the caller holds a required permission, e.g., \texttt{checkCallingPermission()},
leverage \texttt{enforceUriPermission()} to verify whether temporary \texttt{URI}-specific access has been granted~\cite{android-context-api-doc}, 
or rely on ad hoc conditions derived from request fields, caller metadata, or hardcoded values.
This flexibility enables provider-specific authorization, but it can also introduce vulnerabilities when the implemented checks are incomplete, inconsistent, or based on attacker-controlled inputs.

%% file: overview.tex
\section{Overview}
\label{sec:overview}

\subsection{Motivating Example}
\label{sec:motivation}

This section uses a real-world example verified by the vendor to illustrate our research problem, i.e., insufficient access control on sensitive callsites in \texttt{ContentProvider}.
As shown in \autoref{fig:exam2}, this vulnerability resides in \texttt{ContentProviderX}, which inherits from \texttt{ContentProvider} (see line~\ref{line:rms-extends}) within a pre-installed Messaging app shipped on hundreds of millions of devices.
Our vulnerability disclosure communications with the vendor revealed that this component serves as a generic interface for managing the rich communication service (RCS), which delivers services similar to those of the iMessage app in iOS, and it thus provides functions for reading and writing the database. 
While intended to assist database operations, its implementation fails to enforce sufficient access restrictions, \textit{leading to an integrity violation that allows a third-party app to modify SMS or MMS messages on the device without authorization}.\looseness=-1

\input{codes/example2}

Specifically, \texttt{ContentProviderX} is declared in the \texttt{Manifest} file (line~\ref{line:manifest-start} -- line~\ref{line:manifest-end}), where it is specified as accessible to external components (line~\ref{line:manifest-exported}) with a \texttt{READ\_SMS} permission (line~\ref{line:manifest-end}), indicating its restrictions on ``read'' operations.
However, it provides a ``write'' operation in the method \texttt{update} (line~\ref{line:update-start} -- line~\ref{line:update-end}) that ultimately leverages \texttt{SQLiteDatabase.update} (line~\ref{line:db-update}) to modify all SMS and MMS messages stored in the database.
Typically, APIs for database operations are restricted to prevent unauthorized manipulation.
While this \texttt{ContentProviderX.update} method indeed enforces access control, it is limited to checking whether certain business logic is satisfied (see line~\ref{line:check-start} -- line~\ref{line:check-end}), together with a caller-supplied \texttt{URI} parameter through which the caller merely claims to be a sync adapter, without authenticating the identity.
Given the absence of restrictions for the ``write'' operation declared in the \texttt{Manifest}, this interface allows an external component to access and further modify the database.

\input{codes/example2_poc}

As shown in \autoref{lst:rms-poc}, this hypothesized integrity violation has been verified by our attempt to modify the content of an arbitrary SMS via a forged request.
In particular, we first forge an instance of \texttt{ContentValues} and initialize it with a crafted text (line~\ref{line:poc-body}) to overwrite the content of an arbitrary SMS, together with sync-only properties that the provider rejects from non-sync callers (line~\ref{line:poc-deleted} -- line~\ref{line:poc-need-download}).
Next, we forge the sync-adapter identity simply by appending the caller-controlled query parameter \texttt{caller\_is\_syncadapter=true} to the \texttt{URI} at line~\ref{line:poc-uri}.
Since \texttt{callerIsSyncAdapter()} trusts this parameter alone (line~\ref{line:check-adapter-start} -- line~\ref{line:check-adapter-end}), \texttt{isSync} holds and thus short-circuits the check at line~\ref{line:check-start} -- line~\ref{line:check-end}, so the update of these sync-only properties succeeds, and finally we issue our forged malicious update
request by specifying the \texttt{id} of the arbitrary SMS to be modified at line~\ref{line:poc-id}.
This attack indicates that the enforced access control is insufficient to protect the interface, as it lacks identity authentication for a generic API whose malicious intent is buried in this particular context specified by its functionality and argument semantics.

\subsection{Challenges}
\label{sec:challenges}

We have identified three major challenges in addressing the research problem illustrated by the motivating example above. 

\noindent\textbf{C1: How to accurately identify the sensitive callsite.}
The sensitive callsite discussed in this study refers to a system API within an externally accessible \texttt{ContentProvider}, which is internally reachable and feasible from a public interface of the provider.
\texttt{SQLiteDatabase.update} shown in the motivating example (\S\ref{sec:motivation}) is such an example.
It could be used for common and harmless operations if updating sanitized information, but it can also lead to risky operations if it updates private user data, e.g., personal messages.
Unlike APIs chosen as sinks in related taint-analysis-based approaches~\cite{flowdroid, wei2018amandroid, firmscope}, the sensitivity of the API callsites in our study is based not only on their functionality but also on their argument semantics.
Unfortunately, limited attempts have been made to identify such sensitive callsites; they are often excluded due to the potentially unacceptable false positive rate~\cite{firmscope, zhao2020automatic}.
In addition, understanding their argument semantics is a longstanding challenge, as they may be buried within the complex operations of the database, which may also lead to an unacceptable false positive rate~\cite{zhao2020automatic, zuo2019leakscope}.

\noindent\textbf{C2: How to resolve the semantics of enforced access control.}
After identifying the sensitive callsite, the next step is to understand which property and aspect have been enforced with access control.
Specifically, unlike prior work that mainly focuses on reachability or taint-path discovery~\cite{firmscope,zhao2020automatic,flowdroid,wei2018amandroid}, or treats branch conditions as path constraints for feasibility validation~\cite{yang2013appintent,arzt2015using,cadar2013symbolic}, this task requires understanding the purpose of an access control and translating it into formal security properties.
It is challenging because limited attempts have been made to construct those properties, and whether such a translation is feasible remains unknown.

\noindent\textbf{C3: How to reason about and identify insufficient access control.}
After finding sensitive callsites and obtaining their associated access control, we evaluate whether those access restrictions are sufficient.
To this end, a reliable strategy is needed to reason about and justify such an insufficiency.
However, it requires diverse domain knowledge, including a widely recognized standard in related vulnerability identification~\cite{android-severity, samsung-severity, xiaomi-severity}, app-specific context in Android, and reliable semantic reasoning procedures, which are currently missing and remain unknown in the literature.

\subsection{Threat Model and Scope}
\label{sec:threat-model}

\sysname focuses exclusively on the vulnerabilities found in \texttt{ContentProvider}s within pre-installed apps due to insufficient access control. 
Therefore, vulnerable access controls in other Android components like \texttt{service} are out of scope.
This insufficiency may result from the absence of access control, the lack of restrictions on certain operations (e.g., restricting only read operations while leaving write operations unrestricted), or restrictions based on unreliable information, such as identity verification that relies on easily forgeable package names.
In addition, following the line of study for pre-installed apps~\cite{firmscope}, we assume the attackers can only control an unprivileged app, e.g., a third-party app from a public market.
Such an app can only hold permissions that can be obtained without runtime user interaction, i.e., permissions that every app receives automatically at install time.
Moreover, it can interact with a pre-installed app's externally accessible \texttt{ContentProvider} via \texttt{ContentResolver} calls over Binder IPC.

%% file: codes/example2.tex
\begin{figure}[t]
\centering
\begin{tikzpicture}[remember picture]
\node[inner sep=0pt] (codeblock) {%
\begin{minipage}{0.98\columnwidth}
\begin{lstlisting}[language=Java,
    caption={Mismatched checks on a pre-installed Messaging app shipped on hundreds of millions of devices.},
    label={fig:exam2},
    escapeinside={(*@}{@*)},
    basicstyle=\fontsize{6}{7.5}\ttfamily]
<!-- AndroidManifest.xml -->
<provider android:name=".ContentProviderX" (*@\label{line:manifest-start}@*)
          android:authorities="rms_db"
          android:exported="true" (*@\label{line:manifest-exported}@*)
          android:readPermission="android.permission.READ_SMS"/> (*@\label{line:manifest-end}@*)
          
public class ContentProviderX extends ContentProvider { (*@\label{line:rms-extends}@*)
    public int update(Uri uri, ContentValues values, (*@\label{line:update-start}@*)
                      String where, String[] args) {
        SQLiteDatabase db = mOpenHelper.getWritableDatabase();
        boolean isSync   = MmsSmsUtils.callerIsSyncAdapter(uri);
        boolean blocked  = MmsSmsUtils.hasBlockedFlag(uri)
                        || values.containsKey("block_type");
                        
        if (!isSync && !blocked (*@\label{line:check-start}@*)
            && (values.containsKey("is_deleted")
             || values.containsKey("sync_state")
             || values.containsKey("need_download"))) {
            throw new IllegalArgumentException(
                "The non-sync-callers should not specify "
              + "IS_DELETED, SYNC_STATE OR NEED_DOWNLOAD "
              + "in values."); (*@\label{line:check-end}@*)
        }
        
        return db.update("rms_db", values, where, args); (*@\label{line:db-update}@*)
    } (*@\label{line:update-end}@*)
    ...
}

class MmsSmsUtils {
    static final boolean SUPPORT_SYNC_ADAPTER = true;
    
    public static boolean callerIsSyncAdapter(Uri uri) { (*@\label{line:check-adapter-start}@*)
        return SUPPORT_SYNC_ADAPTER && readBooleanQueryParameter(
                    uri, "caller_is_syncadapter", false); (*@\label{line:check-adapter-end}@*)
    }
    
    public static boolean hasBlockedFlag(Uri uri) {
        return readBooleanQueryParameter(
                    uri, "blocked_flag", false);
    }
    ...
}
\end{lstlisting}
\end{minipage}
};
\end{tikzpicture}
\vspace{-0.25in}
\end{figure}

%% file: codes/example2_poc.tex
\begin{figure}[t]
\centering
\begin{minipage}{0.98\columnwidth}
\begin{lstlisting}[language=Java,
    caption={Proof of concept for \autoref{fig:exam2}.},
    label={lst:rms-poc},
    escapeinside={(*@}{@*)},
    basicstyle=\fontsize{6}{7.5}\ttfamily]
ContentValues values = new ContentValues();
values.put("rms_body",      "forged by attacker"); (*@\label{line:poc-body}@*)
values.put("is_deleted",    1); (*@\label{line:poc-deleted}@*)
values.put("sync_state",    0);
values.put("need_download", 1); (*@\label{line:poc-need-download}@*)

Uri uri = Uri.parse("content://rms_db/rms_log"
    + "?caller_is_syncadapter=true"); (*@\label{line:poc-uri}@*)
ctx.getContentResolver().update(uri, values, "_id=43", null); (*@\label{line:poc-id}@*)
\end{lstlisting}
\end{minipage}
\vspace{-0.25in}
\end{figure}

%% file: detailed_design.tex
\section{Detailed Design}
\label{sec:design}

\begin{figure*}[t]
    \centering
    \includegraphics[width=\textwidth]{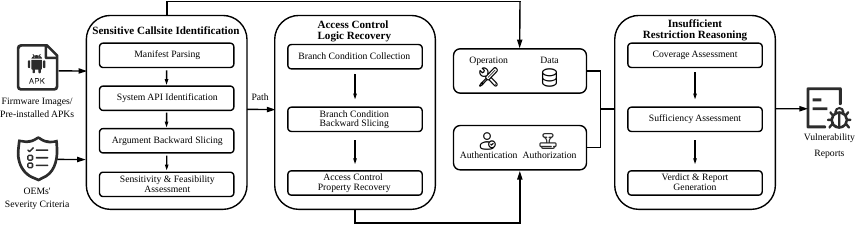}
    \caption{Design overview of \sysname.}
    \label{fig:design_overview}
\end{figure*}

\autoref{fig:design_overview} presents an overview of \sysname, which includes a three-stage pipeline to address the technical challenges summarized in \S\ref{sec:challenges}.
This section explains the details of the implementations following the execution order.
In particular, we first describe how to locate the sensitive callsite in \texttt{ContentProvider}s in \S\ref{sec:design:p1}, then explain how to recover the access control logic to restrict access to these interfaces in \S\ref{sec:design:p2}, and finally present how to reason about insufficient restrictions from the outcome of the previous two procedures in \S\ref{sec:design:p3}.

\subsection{Sensitive Callsite Identification}
\label{sec:design:p1}

The objective of this stage is to identify sensitive callsites within \texttt{ContentProvider}s.
As mentioned in \S\ref{sec:challenges}, a sensitive callsite is a generic system API within an externally accessible provider that could be used for risky functionality depending on its operations and argument semantics.
In addition, there should exist a reachable and feasible path from a public interface of that provider to the generic API.
Since limited attempts have been made to evaluate the sensitivity of our focused callsite, we first propose a novel assessment approach and then present a new suite of techniques to identify all interfaces that meet the required properties.

\input{prompts/prompt_contracts_vertical}

\noindent
\textbf{Callsite sensitivity assessment.}
Motivated by the criteria that Google used to assess the severity of a vulnerability~\cite{android-severity}, we propose to characterize each callsite's sensitivity as a two-label structure.
One label reflects the callsite's functionality category, while the other indicates the category of data involved in the operation.
Specifically, we derive the callsite's functionality from the API's documentation and interpret the involved data based on the argument semantics.

\autoref{tab:sensitive_matrix} summarizes the categories we identified from the official criteria for vulnerability recognition issued by leading Android vendors, including Google~\cite{android-severity}, Samsung~\cite{samsung-severity}, and Xiaomi~\cite{xiaomi-severity}.
In particular, as shown in \autoref{tab:operation_taxonomy}, we classify three operation categories of system APIs serving as the interface in \texttt{ContentProvider}: \textit{($i$)} \textit{configurations} that modify the configuration of security-relevant state, \textit{($ii$)} \textit{function execution} that executes a particular behavior, and 
\textit{($iii$)} \textit{database manipulations} that include basic database operations (e.g., update a table).
Similarly, we classify four privileged and three unprivileged data properties (presented in \autoref{tab:data_context_taxonomy}).
In particular, the privileged properties are \textit{($i$)} device credentials ($P1$), such as TEE secrets and SE keys, \textit{($ii$)} system critical state ($P2$), e.g., verified boot, \textit{($iii$)} system critical settings ($P3$), such as security configuration, and \textit{($iv$)} user privacy ($P4$), such as messages, contacts, and location.
The unprivileged properties, in contrast, cover low-sensitivity data such as the device model.
Together, the operation and data categories provide the semantic context for determining whether a callsite is security-sensitive under the criteria of these vendors~(\autoref{tab:severity_matrix}).\footnote{Two researchers independently open-coded the security-consequence statements in these criteria without a predefined label set, normalized labels with equivalent security semantics, and a third researcher adjudicated the remaining disagreements; the result was frozen before any corpus-level analysis.}

\input{tables/table1}

\noindent
\textbf{Sensitive callsite identification.}
To characterize each callsite's sensitivity, we need to address the functionality of an API and the semantics of its arguments.
While an API's functionality could be derived from its documentation, the argument semantics may be only resolvable from the code.
In addition, we also have to verify the presence of a reachable and feasible path from a public interface of an externally accessible provider to such an API.

Considering that the configuration of a \texttt{ContentProvider} is declared in the \texttt{Manifest} file, \sysname first scans this file to recognize all externally accessible providers by checking whether the attribute \texttt{android:exported} is explicitly set to ``\texttt{true}''.
Next, for each system API within an externally accessible provider, \sysname applies call-graph reachability analysis to verify the presence of a reachable path to such an API.
Meanwhile, it collects all branch conditions in the path to evaluate its feasibility.
Then, \sysname applies backward slicing to resolve the semantics of the API's arguments to characterize its sensitivity according to our newly proposed assessment approach, alongside the API's functionality derived from official documentation.
Finally, it can identify all sensitive callsites by verifying if they meet all required properties.

Since the same callsite may be reached through different paths with different argument values and branch conditions, \sysname analyzes one path at a time.
For every API argument, it performs path-guided backward slicing along the recorded call chain, retaining only the statements that determine the argument along the analyzed path rather than aggregating evidence across all paths reaching the same callsite.
Meanwhile, \sysname collects the branch conditions that control whether execution reaches the callsite. 
For each condition, it records the branch that continues execution toward the callsite and the branch that prevents such execution.
The resulting argument slices and branch conditions provide path-specific evidence for resolving data semantics and path feasibility.

Resolving argument semantics remains challenging because the meaning of an argument may be distributed across multiple statements, methods, and data sources rather than appearing explicitly at the callsite.
Therefore, instead of relying on API-specific heuristic rules, we leverage the LLM, which has been shown to be effective in interpreting program semantics~\cite{ma2023lms, nam2024using}, with a carefully designed prompt to interpret the extracted argument evidence.
Recent studies have shown that LLMs can assist program-analysis reasoning~\cite{wang2024llmdfa,wang2025can}; therefore, we provide the collected branch conditions to the LLM and ask whether they can be jointly satisfied along the analyzed path.

\autoref{fig:prompt-contracts}(a) summarizes the prompt used for callsite sensitivity assessment.
Given an Android system API call and the corresponding argument, the prompt asks the LLM to combine the API signature with backward slices of its arguments.
These slices indicate what resources or data the arguments refer to and where their values come from, following similar approaches~\cite{zhao2020automatic}.
The LLM then maps them to an operation category ($O1$--$O3$) and a data category ($P1$--$P4$ or $U1$--$U3$), and decides whether the callsite is \texttt{sensitive} or \texttt{nonsensitive}.
The output presents the sensitivity assessment, a brief explanation of why it makes such a decision (i.e., reasoning), the identified sensitive execution tuple, and the relevant backward-slice informative elements (i.e., evidence).

\subsection{Access Control Logic Recovery}
\label{sec:design:p2}

Having identified sensitive callsites and obtained the corresponding sensitive paths in \texttt{ContentProvider}s, \sysname next attempts to recover access control mechanisms, i.e., branch conditions, enforced along the path to each interface.
In other words, it recovers the checks on this path.
In particular, we have to resolve the semantics of such branch conditions to uncover the access control purposes and then translate them into security properties that specify their purpose and the information they depend on.
For example, \texttt{<authentication, package\_name>} represents a branch condition to leverage the package name for identity authentication.
Similar to the challenge in the previous task, limited attempts have been made to map branch conditions to access control security properties.
Therefore, following the same idea, we refer to the official criteria mentioned above, summarize a list of security properties, and use an LLM to translate the branches into such properties.\looseness=-1

\autoref{fig:prompt-contracts}(b) summarizes the prompt used for recovering access control mechanisms based on branch conditions on a reachable and feasible path from a public interface of an externally accessible provider to such a callsite.
The prompt asks the LLM to classify the security purpose of each branch condition based on the original condition, the left-hand-side slice, and the right-hand-side slice.
These security purposes include \texttt{authentication} that checks identity, such as UID or signature, \texttt{authorization} that determines whether the caller is allowed to perform an operation, and \texttt{other} if no access control mechanism is identified.
The output includes the classified purpose, structured mechanism, explanation, and supporting slice evidence.\looseness=-1

\subsection{Restriction Context and Sufficiency Reasoning}
\label{sec:design:p3}

The last stage of \sysname is to determine whether the recovered restrictions are sufficient.
In other words, we have to reasonably determine and verify the mismatch between the properties required to secure a specific operation on a specific piece of data and the properties provided by the operation restrictions.
However, the official guideline does not provide concrete rules but rather vague expressions that are extremely difficult to extract deterministic rules from.
Fortunately, an LLM is a great fit here for reasoning about this insufficiency like an expert.
Therefore, we leverage an LLM alongside crafted prompts.

\autoref{fig:prompt-contracts}(c) shows the prompt used for assessing the presence of insufficient protections.
In particular, the LLM is fed with the output from the previous two components and is asked to make decisions.
A protection is considered \texttt{uncovered} if it applies only to an unrelated operation or data; otherwise, it will be \texttt{covered} only if it protects the same sensitive operation and affected data.
In the latter case, the LLM is further asked to decide whether it is \texttt{sufficient}.
A protection is \texttt{sufficient} if it relies on a trusted Android protection mechanism, such as manifest permissions above the normal protection level, resource-matching permission checks, UID checks, or signature checks.
It is \texttt{insufficient} if it only depends on caller-controlled data, checks an unrelated property, or does not enforce real access control~\cite{cwe862, cwe863}.
If no protection covers the sensitive execution, sufficiency is marked as \texttt{not\_applied}.
The output details the coverage verdict, sufficiency verdict, reasoning, sensitive execution, call chain, and evaluated protections.

\subsection{Implementation}
\label{sec:implementation}

\noindent
\textbf{Firmware unpacking and pre-installed app extraction.}
\sysname takes as input the pre-installed apps extracted from a firmware package.
As vendors ship firmware in mutually incompatible containers, we implement a multi-stage extraction procedure driven by magic-number probing.
We first unwrap the outer container, e.g., an A/B OTA package carrying \texttt{payload.bin}, a Samsung \texttt{tar.md5} bundle, an Honor \texttt{UPDATE.APP} archive, or a plain archive of partition images, into a set of partition images.
We then normalize the images: LZ4-compressed payloads are decompressed; chunked and sparse images are converted to raw ones; and a dynamic \texttt{super} image is further split into its constituent logical partitions.
Each raw image is finally identified as EROFS or Ext4 by its superblock magic and unpacked into a file-system tree.
From the trees of the pre-installed partitions, both standard AOSP ones (\texttt{/system}, \texttt{/system\_ext}, \texttt{/product}, \texttt{/vendor}) and vendor-customized ones (\texttt{/mi\_ext}, \texttt{/my\_product}), we collect every APK and framework JAR, additionally unpacking APEX payloads (decompressing \texttt{.capex} when present) to include the APKs and JARs they carry.\looseness=-1

\noindent
\textbf{Program analysis and LLM assistance.}
\sysname is built upon existing program-analysis infrastructure while implementing the task-specific analyses required by our design.
Specifically, we leverage Soot~\cite{soot} for the Jimple intermediate representation and related program-analysis utilities, and use FlowDroid~\cite{flowdroid} solely as a lifecycle-aware call-graph front end.
Unlike prior work that directly relies on generic taint-analysis or slicing frameworks, \sysname implements its own path-specific evidence extraction, including path enumeration from public provider interfaces, argument slicing along individual call paths, and control-dependence analysis for recovering security checks.
These components generate the program evidence consumed by the later LLM-based reasoning stages.
We use DeepSeek~\cite{deepseek} as the default LLM backend and LangChain~\cite{langchain} to orchestrate the prompting and parse the structured JSON outputs.

%% file: prompts/prompt_contracts_vertical.tex
\begin{figure}[t]
\centering
\begingroup
\newcommand{\contractdivider}{\par\vspace{1.5pt}\hrule height 0.35pt\vspace{2pt}}

\begin{tcolorbox}[
  enhanced, sharp corners, boxrule=0.55pt,
  colback=white, colframe=black!65,
  left=4pt,right=4pt,top=2pt,bottom=2pt,
  before skip=0pt,after skip=3pt,
  fontupper=\fontsize{7.4}{8.2}\selectfont,
  title={\scriptsize\textbf{(a) Stage 1: Callsite Sensitivity}},
  coltitle=black,colbacktitle=black!8]
\begin{tabularx}{\linewidth}{@{}>{\bfseries}p{0.11\linewidth}@{\hspace{2pt}}>{\raggedright\arraybackslash}X@{}}
\textsc{Input} & API signature and path-specific argument slices.
\end{tabularx}
\contractdivider
\begin{tabularx}{\linewidth}{@{}>{\bfseries}p{0.11\linewidth}@{\hspace{2pt}}>{\bfseries}p{0.05\linewidth}@{\hspace{2pt}}>{\raggedright\arraybackslash}X@{}}
\textsc{Rule} & 1. & Recover the concrete operation and affected data.\\
 & 2. & Map them to $O1$--$O3$ and $P1$--$P4$/$U1$--$U3$.\\
 & 3. & Accept only when a cited slice supports the decision.
\end{tabularx}
\contractdivider
\textbf{\textsc{Output}}\hfill\textit{structured JSON}\par\smallskip
\noindent\colorbox{black!3}{\parbox{\dimexpr\linewidth-2\fboxsep\relax}{%
\begin{tabularx}{\linewidth}{@{}>{\bfseries}p{0.22\linewidth}@{\hspace{3pt}}>{\raggedright\arraybackslash}X@{}}
Decision & Sensitive or nonsensitive, with rationale.\\
Semantics & Concrete operation $\times$ data with taxonomy labels.\\
Grounding & Cited slice(s) supporting the decision.
\end{tabularx}
\vspace{1pt}\hrule height 0.25pt\vspace{1.5pt}
{\color{black!65}\fontsize{6.5}{7.2}\selectfont
\texttt{fields: sensitivity; reasoning; sensitive\_exec; evidence}}}}
\end{tcolorbox}

\begin{tcolorbox}[
  enhanced, sharp corners, boxrule=0.55pt,
  colback=white, colframe=black!65,
  left=4pt,right=4pt,top=2pt,bottom=2pt,
  before skip=0pt,after skip=3pt,
  fontupper=\fontsize{7.4}{8.2}\selectfont,
  title={\scriptsize\textbf{(b) Stage 2: Check Semantics}},
  coltitle=black,colbacktitle=black!8]
\begin{tabularx}{\linewidth}{@{}>{\bfseries}p{0.11\linewidth}@{\hspace{2pt}}>{\raggedright\arraybackslash}X@{}}
\textsc{Input} & Branch expression and backward slices for both operands.
\end{tabularx}
\contractdivider
\begin{tabularx}{\linewidth}{@{}>{\bfseries}p{0.11\linewidth}@{\hspace{2pt}}>{\bfseries}p{0.05\linewidth}@{\hspace{2pt}}>{\raggedright\arraybackslash}X@{}}
\textsc{Rule} & 1. & Identify both compared objects and their sources.\\
 & 2. & Preserve the exact comparison API or operator.\\
 & 3. & Classify its purpose and cite the supporting slices.
\end{tabularx}
\contractdivider
\textbf{\textsc{Output}}\hfill\textit{structured JSON}\par\smallskip
\noindent\colorbox{black!3}{\parbox{\dimexpr\linewidth-2\fboxsep\relax}{%
\begin{tabularx}{\linewidth}{@{}>{\bfseries}p{0.22\linewidth}@{\hspace{3pt}}>{\raggedright\arraybackslash}X@{}}
Objects & Meaning and source of each compared value.\\
Method & Exact API/operator, e.g., \texttt{equalsIgnoreCase()}, \texttt{equals()}, \texttt{startsWith()}, or \texttt{contains()}.\\
Purpose & Authentication, authorization, or other, with rationale.\\
Grounding & Branch expression and the cited slices of both operands.
\end{tabularx}
\vspace{1pt}\hrule height 0.25pt\vspace{1.5pt}
{\color{black!65}\fontsize{6.5}{7.2}\selectfont
\texttt{fields: purpose; reasoning; mechanism; evidence}}}}
\end{tcolorbox}

\begin{tcolorbox}[
  enhanced, sharp corners, boxrule=0.55pt,
  colback=white, colframe=black!65,
  left=4pt,right=4pt,top=2pt,bottom=2pt,
  before skip=0pt,after skip=0pt,
  fontupper=\fontsize{7.4}{8.2}\selectfont,
  title={\scriptsize\textbf{(c) Stage 3: Protection Sufficiency}},
  coltitle=black,colbacktitle=black!8]
\begin{tabularx}{\linewidth}{@{}>{\bfseries}p{0.11\linewidth}@{\hspace{2pt}}>{\raggedright\arraybackslash}X@{}}
\textsc{Input} & Sensitive execution and call chain, manifest restrictions,
and structured path checks from Stage~2.
\end{tabularx}
\contractdivider
\begin{tabularx}{\linewidth}{@{}>{\bfseries}p{0.11\linewidth}@{\hspace{2pt}}>{\bfseries}p{0.05\linewidth}@{\hspace{2pt}}>{\raggedright\arraybackslash}X@{}}
\textsc{Rule} & 1. & Mark a restriction as covered only if it constrains the Stage~1 operation $\times$ data.\\
 & 2. & If covered, mark it sufficient only if its authority cannot be controlled or forged by the attacker.\\
 & 3. & Aggregate all per-restriction results into the path verdict.
\end{tabularx}
\contractdivider
\textbf{\textsc{Output}}\hfill\textit{structured JSON}\par\smallskip
\noindent\colorbox{black!3}{\parbox{\dimexpr\linewidth-2\fboxsep\relax}{%
\begin{tabularx}{\linewidth}{@{}>{\bfseries}p{0.22\linewidth}@{\hspace{3pt}}>{\raggedright\arraybackslash}X@{}}
Coverage & Covered when at least one restriction constrains the Stage~1 operation $\times$ data.\\
Verdict & (a) \texttt{sufficient}: at least one covering restriction uses authority evidence that the attacker cannot control or forge.\\
 & (b) \texttt{insufficient}: coverage exists, but no covering restriction meets (a).\\
 & (c) \texttt{not\_applied}: no restriction covers.\\
Audit trail & Source, target, authority, and decision for each restriction.
\end{tabularx}
\vspace{1pt}\hrule height 0.25pt\vspace{1.5pt}
{\color{black!65}\fontsize{6.5}{7.2}\selectfont
\texttt{fields: coverage; sufficiency; reasoning; protections}}}}
\end{tcolorbox}

\endgroup
\vspace{0.06in}
\caption{Semantic contracts for the three LLM-assisted stages. Each separates input, rule, human-readable output, and JSON fields; Appendix~\ref{app:prompts} gives the detailed prompts.}
\label{fig:prompt-contracts}
\vspace{-0.25in}
\end{figure}

%% file: tables/table1.tex
\begin{table}[t]
\centering
\scriptsize

\begin{subtable}{\columnwidth}
    \centering
    \begin{tabularx}{\columnwidth}{llX}
    \toprule
    \textbf{} & \textbf{Category} & \textbf{Definition (examples)} \\
    \midrule
    \textbf{$O1$} & Configuration Modification & Modify configuration affecting behavior. \\
    \textbf{$O2$} & Function Execution & Trigger execution (e.g., upload, call). \\
    \textbf{$O3$} & Database Manipulation & Create, read, update, delete. \\
    \bottomrule
    \end{tabularx}
    \caption{Operation categories.}
    \label{tab:operation_taxonomy}
\end{subtable}

\vspace{0.8em}

\begin{subtable}{\columnwidth}
  \centering
  \setlength{\tabcolsep}{4pt}
  \renewcommand{\arraystretch}{1.1}
  \begin{tabularx}{\columnwidth}{@{}l l X@{}}
    \toprule
    \textbf{} & \textbf{Category} & \textbf{Definition (examples)} \\
    \midrule
    \multicolumn{3}{@{}l}{\textit{Privileged data}} \\
    \midrule
    $P1$ & Device Credential &
    Disclosure enables impersonation/irreversible compromise (TEE/SE secrets, auth tokens). \\
    $P2$ & System Critical State &
    OS security-relevant state; integrity-protected even if readable (lock/auth state, verified-boot). \\
    $P3$ & System Critical Setting &
    Persistent config/policy that gates security/authorization (secure settings, allowlists, trust stores). \\
    $P4$ & User Privacy &
    User content/identifiers protected by permission/consent (contacts, location, messages, photos). \\
    \midrule
    \multicolumn{3}{@{}l}{\textit{Unprivileged data}} \\
    \midrule
    $U1$ & Runtime / Telemetry &
    Dynamic low-sensitivity signals commonly available (battery, network type, locale/timezone). \\
    $U2$ & Default / Static Metadata &
    Public constants or device metadata (SDK version, device model, default resources, public endpoints). \\
    $U3$ & Sentinel / Error-as-Data &
    Placeholder/absence/error values that steer logic but carry low sensitivity (\texttt{null}, \texttt{-1}, \texttt{UNKNOWN}, errno). \\
    \bottomrule
  \end{tabularx}
  \caption{Data categories.}
  \label{tab:data_context_taxonomy}
\end{subtable}

\vspace{0.8em}

\begin{subtable}{\columnwidth}
  \centering
  \setlength{\tabcolsep}{2.5pt}
  \renewcommand{\arraystretch}{1.05}
  \begin{tabularx}{\columnwidth}{c *7{>{\centering\arraybackslash}X}}
    \toprule
    \textbf{Oper.\textbackslash Data} &
    \textbf{$P1$} & \textbf{$P2$} & \textbf{$P3$} & \textbf{$P4$} &
    \textbf{$U1$} & \textbf{$U2$} & \textbf{$U3$} \\
    \midrule
    \textbf{$O1$} & \lvlCritical & \lvlCritical & \lvlHigh & \lvlModerate & \lvlModerate & \lvlLow & \lvlIgnore \\
    \textbf{$O2$} & \lvlCritical & \lvlCritical & \lvlHigh & \lvlHigh & \lvlModerate & \lvlIgnore & \lvlIgnore \\
    \textbf{$O3$} & \lvlCritical & \lvlHigh & \lvlModerate & \lvlModerate & \lvlIgnore & \lvlIgnore & \lvlIgnore \\
    \bottomrule
  \end{tabularx}
  \caption{Severity of each operation$\times$data pair.}
  \label{tab:severity_matrix}
\end{subtable}

\vspace{0.8em}
\begin{minipage}{\columnwidth}
\scriptsize
\emph{Note.}
Severity in (c): \lvlCritical~Critical, \lvlHigh~High, \lvlModerate~Moderate, \lvlLow~Low, \lvlIgnore~Ignore.
Privilegedness in (b) is by confidentiality and/or integrity impact; $P2$ may be readable but must not be modified via unprivileged paths.
\end{minipage}

\caption{Taxonomy of categories and the severity carried by each operation$\times$data pair.}
\label{tab:sensitive_matrix}
\end{table}

%% file: evaluation.tex
\section{Evaluation}
\label{evaluation}

In this section, we conduct a large-scale empirical study using \sysname to investigate the prevalence of access control flaws in top-tier Android vendor firmware images, and subsequently evaluate its effectiveness.\footnote{Appendix~\ref{sec:case-study} connects these results to representative vendor-acknowledged cases and working proofs of concept.
}

\subsection{Experiment Setup}
\label{sec:eval:setup}

\noindent
\textbf{Firmware and pre-installed app collection.}
We extract the pre-installed apps from the up-to-date firmware of $10$ influential vendors, including Google and other top vendors of $2025$~\cite{topoem}, and we collect $5$ firmware images from each vendor.
Our primary source is Android Dumps~\cite{androiddumps}, which provides broad vendor coverage, recent images, and build fingerprints.
We also use stock firmware indexed by FirmwareDrive~\cite{firmwaredrive} if a firmware cannot be found in Android Dumps.

\noindent
\textbf{Environment.}
All experiments were conducted on an Ubuntu $20.04$ server with $152$ CPU threads and $1$\,TB RAM.

\noindent
\textbf{LLM configuration.}
    To ensure reproducibility, we set the temperature to $0$ for the DeepSeek and Gemini models and use the default settings for the models and agents from Claude because they do not expose all decoding parameters.

\subsection{Prevalence in the Wild}
\label{sec:eval:aspect}

\input{tables/table_overall_result}

\input{tables/table_corpus}

\autoref{tab:overall-result} presents the overall statistics of our experiment where we apply \sysname to $20{,}006$ pre-installed apps extracted from
$50$ firmware images collected from $10$ Android vendors (see \autoref{tab:eval:corpus}).
After per-vendor deduplication, our corpus contains $1{,}750$ pre-installed apps that have at least one externally accessible \texttt{ContentProvider}.
These apps host a total of $2{,}729$ distinct \texttt{ContentProvider}s, which expose $13{,}131$ public interfaces.
More specifically, starting from these interfaces, \sysname successfully identified $237{,}118$ reachable paths leading to system API invocations.
Following the discovered paths, \sysname extracted $10{,}950$ unique access control branches.
Note that a single branch may check multiple sensitive paths, which accounts for the number of conditions being less than the number of affected paths.
Finally, after pruning paths that are properly protected by access control, \sysname emitted $819$ reachable and feasible vulnerable paths (detailed in \autoref{tab:data-categories}), with the \texttt{deepseek-v3-flash} backend.
In this paper, we treat each vulnerable path as one vulnerability instance, because each path represents a distinct externally reachable execution from a public provider interface to a sensitive callsite under insufficient protection.
In terms of efficiency, preprocessing each firmware image took $16.2$ minutes on average.
For each deduplicated \texttt{ContentProvider}, the analysis took $7.2$ minutes on average, consuming $0.20$M input tokens and $0.06$M output tokens, corresponding to an average cost of US\$$0.12$ when using the DeepSeek model.

\begin{figure}[t]
 \centering
 \includegraphics[width=\columnwidth]{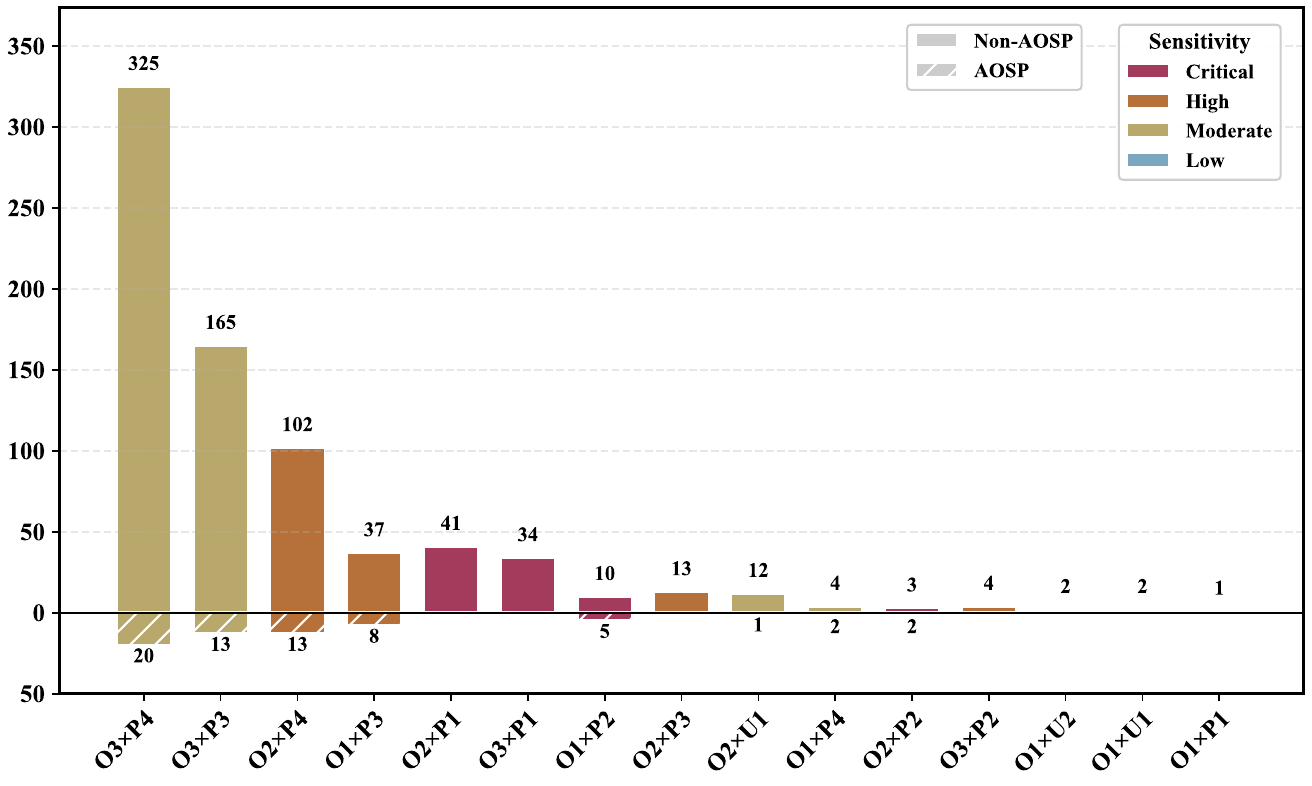}
 \caption{Overall distribution of vulnerable paths.}
 \label{fig:vul_cate_bar}
 \vspace{-0.2in}
\end{figure}

\input{tables/table_checks}

\noindent
\textit{1)} \textbf{Severity analysis:}
Having identified the vulnerable paths, we then characterize their security impacts and present them in \autoref{fig:vul_cate_bar}.
First, we categorize all vulnerable paths into either AOSP or Non-AOSP, which indicates
whether the exposed attack surface is associated primarily with Android's shared platform package set or with components outside the AOSP package set, and we observe that these vulnerable paths substantially come from vendor-specific additions, where they account for $755$ of the $819$ vulnerable paths ($92.2\%$).
Moreover, as mentioned in \S\ref{sec:design:p1}, we characterize the severity in terms of a pair of operation ($O$) and data ($P$).
As shown in \autoref{fig:vul_cate_bar}, these vulnerable paths are highly concentrated in a few particular pairs of the operation and data.
Mapping these pairs through the severity matrix of \autoref{tab:severity_matrix} rates $96$ paths as Critical, $177$ as High, $544$ as Moderate, and $2$ as Low.
These results reveal an alarming situation: the largest group of identified vulnerabilities enables arbitrary manipulation of user-private data through ordinary database operations ($O3\times P4$; e.g., sensitive messaging data and user media files), accounting for $345$ paths, i.e., $42.1\%$.
Beyond user data, the vendor-specific paths also include flaws targeting device credentials such as TEE/SE secrets and authentication tokens.
Collectively, these findings not only underscore the critical severity of \texttt{ContentProvider} misuse in today's Android ecosystem, but also highlight that vendor-provided components represent the primary source of fragile access control implementations.\looseness=-1

\noindent
\textit{2)} \textbf{Access control implementations:}
Having characterized the reachable sensitive operations, we then investigate how these paths are checked in practice.
From the collected sensitive paths, we extract access control restrictions and summarize the results in \autoref{tab:android-access-control-modeling} based on the Android properties they consult.
Specifically, following our threat model (\S\ref{sec:threat-model}), we divide them into secure implementations, such as UID, signature, resource-matching permission, and AppOp checks, and insecure implementations, which solely depend on caller-controlled, contextual, legacy, or compatibility-related properties.
While the results show that many exported interfaces are properly protected by secure access control implementations, a non-negligible number ($2{,}347$ out of $10{,}950$) of checks appear insufficiently secure, where the branch conditions seem security-relevant yet do not actually establish caller authority.
The most common insecure access control pattern is package-name-based checking, which appears $1{,}139$ times, spread across $300$ apps.
Although Android verifies that the package returned by \texttt{getCallingPackage()} belongs to the calling UID, the package-name checks we observe remain bypassable: partial matches (e.g., \texttt{startsWith} or \texttt{contains}) admit attacker-chosen names, and allowlisted packages absent from the device can be claimed by an attacker app.
Such checks therefore do not prove the caller has the system UID, trusted signing identity, or granted capability.
Other insecure access control implementations, including \texttt{SharedUserId}, caller-supplied package fields, intent extras, \texttt{URI} parameters, target SDK versions, build flags, and device-variant checks, can be similarly bypassed via externally controllable interactions.

\noindent
\textit{3)} \textbf{Categories of vulnerable paths:}
\autoref{tab:data-categories} categorizes the $819$ vulnerable paths by the functionality they expose.
The distribution reveals that the problem is not confined to a few conventional privacy APIs.
Instead, vulnerable \texttt{ContentProvider}s have become hidden access points to several major parts of the device experience, including media storage, launcher state, notifications, accessibility-derived context, communication records, and vendor-specific feature control.
The largest groups are user content and UI-related state. 
This is particularly problematic because these providers sit close to daily user interaction: they expose not only stored data such as photos, files, SMS, and call logs, but also behavioral context such as home-screen layout, clipboard content, notification state, recent tasks, and accessibility signals. In other words, the exposed paths can reveal both what users store and how they use the device.
Another interesting observation is that a non-trivial portion of paths reach identity, credential, system, and vendor-specific functionality.
These cases show that vulnerable providers are not merely privacy-leak channels; some of them act as public interfaces to device state, authentication-related data, privileged IPC, secure settings, and vendor feature switches.

\input{tables/table_data_cate}

\subsection{Effectiveness}
 \label{sec:eval:accuracy}

\noindent
{\bf Manual validation.}
To validate the effectiveness of \sysname, we conducted a manual validation on a set of randomly sampled $492$ paths ($60\%$ of the $819$ reported paths) that are distributed in $256$ of the $424$ providers.
In particular, we asked two researchers to independently review every sampled path.
A path was labeled as a true positive only if, under our threat model~(\S\ref{sec:threat-model}), an unprivileged app could reach and execute the sensitive callsite without being blocked by an effective manifest- or code-level restriction.
Under these criteria, the two researchers agreed on $98.2\%$ of the sampled paths and reached consensus on the rest by discussion.
Furthermore, for each vulnerability, one researcher wrote and tested a proof of concept, whose correctness was then double-checked by the other researcher.

\noindent
{\bf Results.}
Finally, $468$ paths across $242$ \texttt{ContentProvider}s were confirmed as true positives and $24$ as false positives, yielding a precision of $95.1\%$ ($95\%$ Wilson confidence interval: $92.8\%$--$96.7\%$).
Seven false positives resulted from access controls enforced asynchronously or across threads that were missed by the static analysis.
The remaining $17$ were LLM-assisted reasoning errors, including nine branch-purpose misclassifications and eight incorrect sufficiency judgments.

\noindent
\textbf{Vendor confirmation.}
We responsibly disclosed the $468$ confirmed vulnerable paths to the affected vendors with working proofs of concept.
At the time of writing, vendors have confirmed $278$ vulnerability reports across $98$ \texttt{ContentProvider}s; some have also been rewarded, patched, or scheduled for remediation.

\noindent
\textbf{Comparison with SOTA.}
To demonstrate the progress made in addressing an overlooked aspect of the state of the art, we compare \sysname with FirmScope~\cite{firmscope} because it focuses on uncovering vulnerabilities that stem from a paradigm similar to ours. 
While no prior work addresses the identical problem in the same scope to enable a perfectly fair comparison, FirmScope, which is the state-of-the-art approach to detecting flaws in pre-installed apps, offers the closest and most meaningful reference point for evaluating our approach.
Since FirmScope is not publicly available, we do not directly run it on our dataset but follow the treatment of unavailable baselines adopted in prior work~\cite{tygr} and only compare within the configuration for which FirmScope was evaluated.

We find that only $13$ of the $468$ paths reach an API in FirmScope's sink set, which upper-bounds the paths FirmScope could report and is a bound on scope overlap. 
This small overlap indicates that \sysname{} uncovers a largely different class of vendor-specific access control flaws, which have been overlooked in the literature.
This gap arises by construction: most of our findings involve generic APIs that FirmScope treats as benign, and \sysname flags them only through semantic reasoning over the callsite.
Directly augmenting FirmScope's sink set with these common APIs, e.g., database and file operations, would make
it unusable because it would indiscriminately flag all \texttt{ContentProvider} operations as vulnerable, which is the reason why prior work excludes these APIs from sink sets.

\looseness=-1

\section{Ablation and Generalizability}
\label{sec:eval:validation}
In this section, we carry out an ablation study on a manually validated dataset, to validate the effectiveness of each key component of \sysname.
Then, we further evaluate how well \sysname generalizes across backend models, prompt variants, and commercial general-purpose coding agents.

\noindent

\noindent\textbf{Dataset.}
We manually built a dataset of $500$ code paths ($250$ vulnerable and $250$ non-vulnerable) by decompiling randomly selected providers from the $2{,}729$ externally accessible ones with JADX.
Two analysts with research experience in mobile security independently reviewed the \texttt{ContentProvider} code and, for every path, attempted to construct and run a working proof of concept on our own devices.
Because each judgment rests on an executed attempt rather than on code reading alone, the two analysts agreed on $497$ of the $500$ paths, and resolved the remaining three by discussion under the criteria of \S\ref{sec:design:p1}.

\input{tables/table_ablation.tex}

\subsection{Ablation Analysis}
We evaluated four configurations that sequentially incorporate \sysname's core design components---i.e., the static code slicing, the taxonomy guidance, and the sufficiency criteria---building incrementally toward the complete tool.
The results are shown in the first four rows of \autoref{tab:ablation-study}, all obtained with the \texttt{deepseek-v3-flash} backend used in the corpus-scale analysis.
Note that the configurations differ only in the intermediate states produced by the active design components, while the ultimate objective---determining whether a given path is vulnerable---and the backend LLM are held constant across all settings. Specifically:

\noindent\textbf{C0: No extracted evidence.}
    As the starting point, we give the backend only the analysis task and the decompiled code, withholding all evidence produced by \sysname's static analysis.
    This configuration performs close to chance, at $51.7\%$ F1 and $50.6\%$ accuracy, while costing US\$$17.27$, over $13\times$ the cost of the full framework, because every judgment triggers its own code retrieval.
    The extracted path, slice, and manifest evidence is therefore a prerequisite for the reasoning stages that follow, rather than a convenience.

\noindent\textbf{C1: Code slices evidence only (E).}
    When provided with the sliced code and manifest information of a path, the backend LLM can concentrate more on the target code compared to \textbf{C0}; however, the tool flags nearly every reachable path as vulnerable.
    While it successfully captures almost all vulnerable instances, it also misclassifies $206$ of the $250$ non-vulnerable paths as vulnerable, yielding high recall but impractically low precision.
    This result confirms that, without a notion of what makes a callsite sensitive, almost anything reachable could be considered as vulnerable.

\noindent\textbf{C2: C1 with taxonomy guidance (T).}
    Compared with \textbf{C1} in \autoref{tab:ablation-study}, \textbf{C2} sharply raises precision to $94.6\%$ once the taxonomies are added, effectively reducing the false positives.
    However, recall drops to $56.4\%$, as \textbf{C2} can identify the sensitive operation and the data it touches but lacks a criterion for judging whether a security check provides sufficient protection.
    Consequently, it treats the presence of any check as adequate and labels it as safe, leading to a substantial amount of misjudgment on those paths whose checks are present but forgeable, increasing the false negatives.

\noindent\textbf{C3: C2 with sufficiency reasoning (S).}
    According to the results of \textbf{C2} and \textbf{C3} in \autoref{tab:ablation-study}, incorporating sufficiency reasoning recovers those previously misclassified paths, lifting recall to $92.0\%$ and F1 to $92.6\%$ while reducing misclassified non-vulnerable paths to only $17$ of the $250$, down from $206$ at \textbf{C1}.
    This demonstrates that the sufficiency reasoning effectively distinguishes between security-relevant checks and truly protective ones, which substantially reduces false negatives.

\subsection{Evaluation of Generalizability}
We evaluate whether \sysname remains effective across different backend models and prompt variants, and whether its core design components also improve commercial general-purpose coding agents.

\noindent
\textbf{Model sensitivity.}
In addition to the ablation study above, we further run the complete \sysname framework on three additional backends, \texttt{deepseek-v4-pro}, \texttt{gemini-3.1-pro}, and \texttt{claude-opus-5}, varying only the backend.
Across the four backends, precision varies by at most $1.9$ percentage points, recall by $4.8$, and accuracy by $2.2$, with accuracy remaining at or above $92.4\%$.
Notably, \texttt{gemini-3.1-pro} and \texttt{claude-opus-5} improve accuracy over \texttt{deepseek-v3-flash} by only $1.4$ and $2.0$ percentage points, respectively, despite costing $68\times$ and $139\times$ as much.
These results indicate that \sysname's effectiveness is not tied to a particular backend and that substantially higher model costs yield only marginal accuracy gains.\looseness=-1

\noindent
\textbf{Prompt sensitivity.}
Beyond the backend, a second concern is whether the results depend on how the prompts happen to be worded.
We therefore construct $20$ semantically equivalent variants of \sysname's three-stage prompts, obtained by paraphrasing the instructions and reordering their independent parts while keeping the input evidence, the taxonomy definitions, the decision criteria, and the JSON output schema unchanged; two authors reviewed every variant to confirm that no decision rule was added, removed, or reweighted.
We then rerun the full pipeline on the same $500$ paths for each variant, holding the backend fixed at \texttt{deepseek-v3-flash}.
Accuracy ranges from $92.2\%$ to $92.8\%$ (median $92.6\%$) and F1 from $92.1\%$ to $92.9\%$, a spread of three of the $500$ paths, and no variant differs from the original prompt on more than nine path-level verdicts.
The results are therefore driven by the extracted evidence and the decision criteria rather than by the particular phrasing of the prompts.

\noindent
\textbf{Comparison with a commercial code agent.}
We then evaluate how the proposed \sysname performs when compared with the advanced commercial code agent, i.e., Claude Code.
Two complementary experiments are conducted:
\begin{itemize}[leftmargin=*]
    \item \textbf{Same backend LLM model, different analysis formulations:} To distinguish between differences caused by analysis formulation and those caused by model capability, we replace the backend LLM in \sysname with the one integrated within Claude Code, i.e., \texttt{claude-opus-5}, and keep all other inputs identical, i.e., the provided decompiled code, target \texttt{ContentProvider}, and task specifications (e.g., threat model).\looseness=-1

    \textbf{Result.} The comparative results are shown in the lines identified by \textbf{\sysname-C3 (\texttt{claude-opus-5})} and \textbf{Agent-C0 (\texttt{claude-opus-5})} in \autoref{tab:ablation-study}. 
    It is observed that \sysname with \texttt{claude-opus-5} backend significantly surpasses its agent counterpart in all four metrics.
    Although the Claude Code baseline detects some vulnerabilities, it both misses vulnerable paths and flags many safe ones due to omitting critical paths, mischaracterizing data sensitivity, and incorrectly judging the sufficiency of protection checks.
    These findings confirm that the analysis formulation---rather than mere model capacity---is the primary driver of detection performance.

    \item \textbf{Same agent baseline, different input contexts:} To isolate the contribution of \sysname's design, we start from an LLM agent given only the decompiled code and the task definition, which are similar to the previous experiment, and then sequentially enrich the input contexts provided to the agent with intermediate results from \sysname's components, i.e., the code slice evidence, the taxonomy guidance, and the sufficiency reasoning.
    We adopt the Claude Code agent with different backend LLMs (i.e., \texttt{claude-sonnet-5} and \texttt{claude-opus-5}) for comprehensiveness.

    \textbf{Result.} The results are presented in the lower half of \autoref{tab:ablation-study}.
    Starting from the initial configuration, we observe an overall upward trend in the metrics as we sequentially adopt the input contexts used in \sysname, confirming the positive contribution of each design stage. Compared with the other two components, the taxonomy guidance yields relatively modest gains.
    We attribute this to the agents' inherent proficiency in identifying sensitive data, which suggests that their training may have imparted a data-sensitivity characterization capability akin to that explicitly encoded in our taxonomy---an observation consistent with recent findings~\cite{carlini2026zerodays}.
    We also note that, unlike \sysname, which directly employs an LLM as its backend, the code agent is more susceptible to distraction when processing complex input instructions.
    This manifests as fluctuations in individual metrics (e.g., recall) and leads to inferior overall performance relative to \sysname with the same backend.
    Nevertheless, the experimental results confirm that even powerful LLM agents benefit from our solutions.

\end{itemize}

%% file: tables/table_overall_result.tex
\begin{table}[t]
\centering
\small
\setlength{\tabcolsep}{4pt}
\renewcommand{\arraystretch}{1.08}

\begin{tabularx}{\columnwidth}{@{}X r@{}}
\toprule
\textbf{Metric} & \textbf{Value} \\
\midrule

\multicolumn{2}{@{}l}{\textit{Corpus and exposed interfaces}} \\
$\Sigma$ \# Firmware images & $50$ \\
$\Sigma$ \# Pre-installed apps & $20{,}006$ \\
$\Sigma$ \# Dedup. apps w/ ext. \texttt{ContentProvider}s & $1{,}750$ \\
$\Sigma$ \# Externally accessible \texttt{ContentProvider}s & $2{,}729$ \\
$\Sigma$ \# Public interfaces & $13{,}131$ \\

\addlinespace[2pt]
\midrule
\multicolumn{2}{@{}l}{\textit{Path and access control analysis}} \\
$\Sigma$ \# Reachable paths & $237{,}118$ \\
$\Sigma$ \# Access control branch conditions & $10{,}950$ \\

\addlinespace[2pt]
\midrule
\multicolumn{2}{@{}l}{\textit{Vulnerability findings}} \\
$\Sigma$ \# Reported paths & $819$ \\
$\Sigma$ \# Affected \texttt{ContentProvider}s & $424$ \\
$\Sigma$ \# Affected apps & $197$ \\
$\Sigma$ \# Affected vendors & $10$ \\

\addlinespace[2pt]
\midrule
\multicolumn{2}{@{}l}{\textit{Efficiency and cost}} \\
\# Avg. preprocessing time per firmware & $16.2$ min \\
\# Avg. analysis time per provider & $7.2$ min \\
\# Avg. LLM tokens per provider (in/out) & $0.20$M / $0.06$M \\
\# Avg. LLM cost per provider (DeepSeek) & US\$$0.12$ \\

\bottomrule
\end{tabularx}

\begin{minipage}{0.98\columnwidth}
\footnotesize
\end{minipage}
\caption{Overall experimental results.}
\label{tab:overall-result}
\end{table}

%% file: tables/table_corpus.tex
\begin{table*}[t]
\centering
\small
\setlength{\tabcolsep}{4pt}
\renewcommand{\arraystretch}{1.08}

\begin{tabular*}{\textwidth}{@{\extracolsep{\fill}}lrrrrrrrrr@{}}
\toprule
\textbf{Vendor} & \textbf{\#FWs}
& \multicolumn{2}{c}{\textbf{\#Pre-installed Apps}}
& \multicolumn{3}{c}{\textbf{\#Dedup. Apps w/ Ext. CPs}}
& \textbf{\#Dedup.}
& \textbf{\#Public}
& \textbf{\#Reachable} \\
\cmidrule(lr){3-4} \cmidrule(lr){5-7}
&
& \textbf{All}
& \textbf{w/ Ext. CPs}
& \textbf{All}
& \textbf{AOSP}
& \textbf{Non-AOSP}
& \textbf{CPs}
& \textbf{Interfaces}
& \textbf{Paths} \\
\midrule
Google & $5$ & $1{,}725$ & $192$ & $103$ & $25$ & $78$ & $150$ & $721$ & $14{,}161$ \\
Samsung & $5$ & $2{,}430$ & $660$ & $209$ & $22$ & $187$ & $401$ & $2{,}003$ & $21{,}080$ \\
Xiaomi & $5$ & $2{,}199$ & $471$ & $219$ & $34$ & $185$ & $440$ & $1{,}902$ & $23{,}457$ \\
OPPO & $5$ & $2{,}088$ & $371$ & $175$ & $25$ & $150$ & $297$ & $1{,}617$ & $34{,}451$ \\
vivo & $5$ & $1{,}700$ & $270$ & $143$ & $26$ & $117$ & $190$ & $1{,}018$ & $22{,}880$ \\
realme & $5$ & $2{,}123$ & $451$ & $209$ & $29$ & $180$ & $357$ & $1{,}631$ & $27{,}998$ \\
Honor & $5$ & $2{,}059$ & $279$ & $149$ & $25$ & $124$ & $212$ & $1{,}077$ & $25{,}068$ \\
Infinix & $5$ & $1{,}894$ & $375$ & $167$ & $30$ & $137$ & $188$ & $879$ & $23{,}355$ \\
Tecno & $5$ & $1{,}482$ & $294$ & $216$ & $32$ & $184$ & $255$ & $1{,}008$ & $19{,}349$ \\
Motorola & $5$ & $2{,}306$ & $318$ & $160$ & $26$ & $134$ & $239$ & $1{,}275$ & $25{,}319$ \\
\midrule
\textbf{Total} & $\textbf{50}$ & $\textbf{20{,}006}$ & $\textbf{3{,}681}$ & $\textbf{1{,}750}$ & $\textbf{274}$ & $\textbf{1{,}476}$ & $\textbf{2{,}729}$ & $\textbf{13{,}131}$ & $\textbf{237{,}118}$ \\
\bottomrule
\end{tabular*}

\vspace{3pt}
\begin{minipage}{\textwidth}
\footnotesize
\textit{Note.}
FW denotes firmware image; CP denotes \texttt{ContentProvider};
Ext. CP denotes an externally accessible \texttt{ContentProvider}.
Deduplication is performed per vendor. Apps are deduplicated by package
name, and CPs are deduplicated by package and component name.
\end{minipage}

\caption{Per-vendor app and \texttt{ContentProvider} statistics.}
\label{tab:eval:corpus}
\end{table*}

%% file: tables/table_checks.tex
\begin{table*}[t]
  \centering

  \footnotesize
  \setlength{\tabcolsep}{3.2pt}
  \renewcommand{\arraystretch}{1.12}

  \begin{tabularx}{\textwidth}{p{0.008\textwidth} | p{0.08\textwidth} L{0.16\textwidth} X r r}
    \hline
    \textbf{} & \textbf{Status} & \textbf{Property} & \textbf{Typical Use / Modeling Rationale} 
    & \textbf{\#Apps} & \textbf{\#Checks} \\
    \hline

    \hline
      \multirow{12}{*}{\rotatebox[origin=c]{90}{\textbf{Authentication}}}
        & \textbf{Secure}
        & UID
        & Direct Binder UID identity check, e.g., \texttt{Binder.getCallingUid()}.
        & 234 & 1,028 \\
      
      &
        & Signature
        & Package signature validation check, e.g., \texttt{checkSignatures(uid1, uid2)}.
        & 285 & 2,081 \\
      
      &
        & System Flag
        & System-app flag based trust check, e.g., \texttt{applicationInfo.flags}.
        & 137 & 776 \\
      
      &
        & Same Process
        & Binder UID self-call consistency check, e.g., \texttt{Process.myUid()}.
        & 160 & 1,048 \\

    \cline{2-6}

      &
        \textbf{Insecure}
        & Package Name
        & Partial or squattable package-name match, e.g., \texttt{contains, startsWith}.
        & 300 & 1,139 \\
      
      &
        & Shared User ID
        & Shared UID attribution check, e.g., \texttt{packageManager.getNameForUid()}.
        & 1 & 2 \\
      
      &
        & Caller-Supplied Package
        & Caller-claimed package name, e.g., \texttt{intent.getStringExtra("pkgName")}.
        & 36 & 42 \\
      
      &
        & Self-Claimed Extra
        & Caller-input intent or bundle value, e.g., \texttt{extras.getBoolean("from\_xxx")}.
        & 81 & 200 \\
      
      &
        & URI Parameter
        & Caller-input URI identity flag, e.g., \texttt{uri.readQueryParam("caller\_is\_xxx")}.
        & 136 & 439 \\
      
      &
        & Current User ID
        & Active-user state used as caller identity, e.g., \texttt{getCurrentUser()}.
        & 28 & 40 \\
      
      &
        & Hardcoded Token
        & Static recoverable token-based check, e.g., \texttt{token.equals("xxxxxxxx")}.
        & 35 & 57 \\
      
      &
        & Weak-Random Token
        & Low-entropy token-based identity check, e.g., derived from a few address bits.
        & 1 & 1 \\

    \hline

    \hline
      \multirow{5}{*}{\rotatebox[origin=c]{90}{\textbf{Authorization}}}
        & \textbf{Secure}
        & Consistent Permission
        & Resource-matching permission enforcement, e.g., \texttt{READ/WRITE\_SMS} for SMS data.
        & 443 & 3,629 \\
      
      &
        & AppOp
        & Operation-aligned AppOps enforcement, e.g., \texttt{setAppOps} for relevant permissions.
        & 27 & 41 \\

    \cline{2-6}

      &
        \textbf{Insecure}
        & Inconsistent Permission
        & Resource-mismatched permission enforcement, e.g., a weather permission for GPS.
        & 7 & 56 \\
      
      &
        & Self Permission Check
        & Self-UID checks are used to restrict external calls, e.g., \texttt{checkSelfPermission()}.
        & 26 & 44 \\
      
      &
        & Intent Action / Category
        & Intent metadata used as access gate, e.g., action strings protecting privileges.
        & 27 & 41 \\

    \hline

    \hline
      \multirow{4}{*}{\rotatebox[origin=c]{90}{\textbf{Other}}}
        & \textbf{Insecure}
        & Target SDK Version
        & Compatibility metadata used as a security gate, e.g., \texttt{targetSdkVersion < N}.
        & 70 & 142 \\
      
      &
        & Region
        & Regional build flag-based gate, e.g., \texttt{Build.IS\_INTERNATIONAL\_BUILD}.
        & 20 & 29 \\
      
      &
        & Device Variant
        & Device feature flag-based gate, e.g., \texttt{SettingsFeatures.isFlipDevice}.
        & 6 & 10 \\
      
      &
        & Build Type
        & Build flavor used as security gate, e.g., \texttt{Build.TYPE.equals("user")}.
        & 71 & 105 \\

    \hline
  \end{tabularx}
  \caption{Android properties used in dynamic access control checks.}
  \label{tab:android-access-control-modeling}
  \vspace{-0.1in}
\end{table*}

%% file: tables/table_data_cate.tex
\begin{table*}[t]
\centering

\footnotesize
\setlength{\tabcolsep}{3.5pt}
\renewcommand{\arraystretch}{1.08}

\begin{tabularx}{\textwidth}{
@{}
L{0.145\textwidth}
L{0.175\textwidth}
>{\raggedleft\arraybackslash}p{0.055\textwidth}
>{\raggedleft\arraybackslash}p{0.055\textwidth}
Y
@{}}
\toprule
\textbf{Group} &
\textbf{Subcategory} &
\textbf{\#Apps} &
\textbf{\#Paths} &
\textbf{Representative data} \\
\midrule

\multirow[t]{4}{=}{\group{Communication}}
& SMS / MMS messages & $4$ & $48$
& Message bodies, sender addresses, attachments, quick replies, and read states. \\

& Call logs & $9$ & $25$
& Call records with numbers, durations, timestamps, and rejection replies. \\

& SIM \& carrier state & $6$ & $28$
& SIM records, carrier settings, and private network configuration data. \\

& Contacts \& address book & $5$ & $27$
& Contact records with phone numbers, emails, profiles, and mappings. \\

\addlinespace[2pt]
\midrule

\multirow[t]{4}{=}{\group{Identity \& Credentials}}
& Biometric data & $3$ & $4$
& Face-service controls and face-cluster names associated with photos. \\

& Crypto keys \& material & $3$ & $10$
& Bluetooth keys, account keys, security tokens, and sharing keys. \\

& Account \& auth tokens & $14$ & $37$
& Account states, push identifiers, and vendor account-binding keys. \\

& Device \& advertising IDs & $3$ & $7$
& Advertising identifiers, vendor user identifiers, and diagnostic device IDs. \\

\addlinespace[2pt]
\midrule

\multirow[t]{5}{=}{\group{User Content \& Media}}
& Browser data & $11$ & $32$
& Browser homepage settings and writable configuration affecting navigation. \\

& Photos, videos \& gallery & $30$ & $118$
& Source images, media metadata, location tags, and video viewing history. \\

& Wallpapers \& stickers & $13$ & $29$
& Lock-screen wallpaper settings, theme records, and usage history. \\

& File exposure & $21$ & $48$
& Log dumps, clipboard caches, and cross-device transfer metadata. \\

& Downloads \& OTA & $8$ & $24$
& Update metadata, download states, and installation activation records. \\

\addlinespace[2pt]
\midrule

\multirow[t]{2}{=}{\group{Sensors \& Location}}
& Health \& fitness & $5$ & $10$
& Health analysis records and menstruation prediction information. \\

& Location \& weather & $6$ & $22$
& GPS coordinates, location history, city lists, and weather locations. \\

\addlinespace[2pt]
\midrule

\multirow[t]{6}{=}{\group{UI \& Accessibility}}
& Clipboard \& drag & $6$ & $21$
& Clipboard contents, drag-transfer entries, and input-method phrase lists. \\

& App usage \& recent tasks & $4$ & $8$
& Booking records, viewing behavior, and daily application usage time. \\

& Accessibility services & $6$ & $44$
& Accessibility-derived foreground application and in-app state information. \\

& Launcher \& home screen & $17$ & $61$
& Home layouts, installed applications, hidden apps, and whitelist records. \\

& Notifications \& alerts & $9$ & $41$
& Update banners, notification badges, and scheduled power-state alerts. \\

& Lock screen \& screen state & $4$ & $5$
& Screen-recorder status and remote-controller device state information. \\

\addlinespace[2pt]
\midrule

\multirow[t]{2}{=}{\group{Connectivity}}
& Wi-Fi \& Bluetooth & $3$ & $19$
& Bluetooth pairing records, MAC addresses, and wireless configuration data. \\

& IoT \& smart-home pairing & $3$ & $5$
& Paired vehicle devices and smart-home Bluetooth pairing records. \\

\addlinespace[2pt]
\midrule

\multirow[t]{5}{=}{\group{System \& Platform}}
& Privileged IPC \& URI & $13$ & $18$
& Agent command channels and trusted execution interfaces. \\

& Logs \& diagnostics & $11$ & $19$
& Diagnostic logs from assistants, cameras, messages, and launchers. \\

& Settings values & $9$ & $15$
& Writable secure settings and global configuration tables. \\

& Device admin \& power & $4$ & $15$
& Parental-control flags, passcode states, batteries, and power schedules. \\

& App/package management & $6$ & $13$
& Installation caches, package state tables, and installed application lists. \\

\addlinespace[2pt]
\midrule

\multirow[t]{4}{=}{\group{Vendor / Apps}}
& Games \& game mode & $3$ & $13$
& Game acceleration states, VPN switches, and location-tagged game data. \\

& Payment \& wallet records & $1$ & $1$
& Biometric payment switches and wallet-related configuration states. \\

& On-device AI \& assistants & $5$ & $20$
& AI action surfaces and assistant inputs for personal context. \\

& OEM-specific config & $14$ & $32$
& NFC states, split-screen settings, push toggles, and camera tasks. \\

\midrule
\multicolumn{2}{r}{\textbf{Total}} &
$\textbf{197}$ &
$\textbf{819}$ &
--- \\
\bottomrule
\end{tabularx}
\begin{minipage}{\textwidth}
\footnotesize
\vspace{3pt}
\textit{Note.}
\#Paths is additive; \#Apps is deduplicated at the total row and may overlap across categories.
\end{minipage}
\caption{Categories of vulnerable paths discovered across the analyzed firmware images.}
\label{tab:data-categories}
\vspace{-0.15in}
\end{table*}

%% file: tables/table_ablation.tex
\begin{table*}[t]
\centering
\small
\setlength{\tabcolsep}{5pt}
\setlength{\aboverulesep}{0pt}
\setlength{\belowrulesep}{0pt}
\setlength{\extrarowheight}{.55ex}
\renewcommand{\arraystretch}{1.0}
\begin{tabular*}{\textwidth}{@{\extracolsep{\fill}}c|cccrrrrrrrrr@{}}
\toprule
& \multicolumn{2}{c}{\textbf{Config.}}
& \multirow{2}{*}{\textbf{Model}}
& \multicolumn{4}{c}{\textbf{Counts}}
& \multicolumn{4}{c}{\textbf{Scores (\%)}}
& \textbf{Cost} \\
\cmidrule(lr){2-3}\cmidrule(lr){5-8}\cmidrule(lr){9-12}
& \textbf{ID} & \lvlhead & & \textbf{TP} & \textbf{TN} & \textbf{FP} & \textbf{FN}
& \textbf{Prec.} & \textbf{Rec.} & \textbf{F1} & \textbf{Acc.} & \textbf{(US\$)} \\
\midrule
\multirow{7}{*}{\rotatebox[origin=c]{90}{\textbf{\sysname}}}
& C0 & \lvl{0} & \texttt{deepseek-v3-flash} & $132$ & $121$ &  $129$ &  $118$ & $50.6$ & $52.8$ & $51.7$ & $50.6$ & $17.27$ \\
& C1 & \lvl{1} & \texttt{deepseek-v3-flash} & $232$ & $44$ & $206$ & $18$ & $53.0$ & $92.8$ & $67.4$ & $55.2$ & $0.82$ \\
& C2 & \lvl{2} & \texttt{deepseek-v3-flash} & $141$ & $242$ & $8$ & $109$ & $\mathbf{94.6}$ & $56.4$ & $70.7$ & $76.6$ & $0.98$ \\
& C3 & \lvl{3} & \texttt{deepseek-v3-flash} & $230$ & $233$ & $17$ & $20$ & $93.1$ & $92.0$ & $92.6$ & $92.6$ & $1.29$ \\
& C3 & \lvl{3} & \texttt{deepseek-v4-pro} & $234$ & $228$ & $22$ & $16$ & $91.4$ & $93.6$ & $92.5$ & $92.4$ & $6.91$ \\
& C3 & \lvl{3} & \texttt{gemini-3.1-pro} & $237$ & $233$ & $17$ & $13$ & $93.3$ & $94.8$ & $94.0$ & $94.0$ & $88.32$ \\
& C3 & \lvl{3} & \texttt{claude-opus-5} & $242$ & $231$ & $19$ & $8$ & $92.7$ & $\mathbf{96.8}$ & $\mathbf{94.7}$ & $\mathbf{94.6}$ & $178.67$ \\
\midrule
\multirow{8}{*}{\rotatebox[origin=c]{90}{\textbf{Agent}}}
& C0 & \lvl{0} & \texttt{claude-sonnet-5} & $198$ & $102$ & $148$ & $52$ & $57.2$ & $79.2$ & $66.4$ & $60.0$ & $297.25$ \\
& C1 & \lvl{1} & \texttt{claude-sonnet-5} & $186$ & $212$ & $38$ & $64$ & $83.0$ & $74.4$ & $78.5$ & $79.6$ & $255.93$ \\
& C2 & \lvl{2} & \texttt{claude-sonnet-5} & $183$ & $222$ & $28$ & $67$ & $86.7$ & $73.2$ & $79.4$ & $81.0$ & $283.67$ \\
& C3 & \lvl{3} & \texttt{claude-sonnet-5} & $215$ & $212$ & $38$ & $35$ & $85.0$ & $86.0$ & $85.5$ & $85.4$ & $254.84$ \\
& C0 & \lvl{0} & \texttt{claude-opus-5} & $206$ & $114$ & $136$ & $44$ & $60.2$ & $82.4$ & $69.6$ & $64.0$ & $548.79$ \\
& C1 & \lvl{1} & \texttt{claude-opus-5} & $162$ & $232$ & $18$ & $88$ & $90.0$ & $64.8$ & $75.3$ & $78.8$ & $449.21$ \\
& C2 & \lvl{2} & \texttt{claude-opus-5} & $161$ & $239$ & $11$ & $89$ & $\mathbf{93.6}$ & $64.4$ & $76.3$ & $80.0$ & $407.42$ \\
& C3 & \lvl{3} & \texttt{claude-opus-5} & $222$ & $218$ & $32$ & $28$ & $87.4$ & $\mathbf{88.8}$ & $\mathbf{88.1}$ & $\mathbf{88.0}$ & $446.85$ \\
\bottomrule
\end{tabular*}

\caption{Component ablation, model sensitivity, and comparison with a commercial code agent.}
\label{tab:ablation-study}
\vspace{-0.15in}
\end{table*}

%% file: discussion.tex
\section{Discussion}

\subsection{Limitations}
\label{sec:limitations}

While \sysname has identified many new vulnerabilities, it is not perfect.
First, it inherits limitations from those static analysis components we built upon.
These limitations include, but are not limited to, the inability to \textit{($i$)} track dataflows through certain components, \textit{($ii$)} handle sophisticated obfuscation such as reflection, and \textit{($iii$)} analyze native libraries.
In addition, while we carefully crafted prompts to guide the LLM to work properly, \sysname still suffers from hallucination problems, e.g., misinterpretation and misreasoning, leading to false positives.\looseness=-1

\subsection{Taxonomy Coverage}

We conducted an open-category audit on all $500$ paths in the dataset presented in \S\ref{sec:eval:validation}, none of which was used to develop the taxonomies or the prompts.
For the data and operation taxonomy, the annotators described each path's operation and affected data in free-form terms, mapped each description to the frozen taxonomy, and recorded whether a new category was required.
All $500$ descriptions were assigned to an existing category, confirming the coverage of $100\%$.
Unlike the data and operation taxonomies, the branch-purpose taxonomy avoids the open-category problem because \texttt{other} is exhaustive by construction: every check that does not enforce authentication or authorization is retained under this category.\looseness=-1

Although we did not discover any exceptions, we acknowledge that the established data and operation taxonomies may have blind spots: a previously unseen resource type, or a security distinction finer than the current categories allow, could fall outside the taxonomy and cause \sysname to miss the corresponding callsites.
Nevertheless, extending the taxonomies is a straightforward one-time effort that requires no changes to the pipeline or prompt templates; we therefore believe \sysname could be readily adapted to maintain satisfactory performance even in the presence of such previously unseen cases.

\subsection{Broader Applicability}
Although \sysname is implemented and evaluated for Android \texttt{ContentProvider}s, its underlying reasoning pattern is more general: it determines callsite sensitivity from the semantics of both the operation and the resource involved, and then evaluates whether the surrounding checks establish sufficient authority under the threat model. This reasoning may also apply to systems that expose generic database, file, storage, or configuration interfaces, where API identity and the mere presence of an access control check may not reveal the actual security risk. Applying \sysname to another platform, however, would require a platform-specific analysis frontend, operation and resource taxonomies, a corresponding threat model, and rules defining sufficient authorization. We leave such extensions and their empirical evaluation to future work.

\subsection{Mitigation}
\label{sec:mitigation}

\noindent
\textbf{Avoiding overly powerful interfaces.}
In this study, we discovered pre-installed apps that utilize generic and powerful interfaces to deliver functionality that could also be achieved with less powerful APIs.
One typical pattern in our vulnerable cases is the widespread use of \texttt{ContentProvider.call()}, which is designed for provider-defined operations that do not naturally fit the standard table-like CRUD model~\cite{aospprovider}.
We observe that vendor providers often extend this flexible interface into a general-purpose command channel for privileged functionality, while the Android framework does not enforce fine-grained permission checks on this method, making security failures at these interfaces particularly impactful.
Therefore, developers should avoid exposing privileged operations through overly generic interfaces when more specialized and restricted APIs, e.g., the standard CRUD APIs, are enough.

\noindent
\textbf{Using trusted authority information for identity verification.}
Our analysis reveals widespread reliance on weak or caller-controllable identities, including string-level identities such as package names, shared UID names, and request-provided identifiers, as authorization boundaries.
These values do not necessarily establish the authority of the caller and may be forged or misused by unprivileged applications.
Therefore, sensitive IPC interfaces should enforce permissions that match the protected functionality, e.g., SMS interfaces should require SMS permissions.
For vendor-private functionality, the callee should further verify the real Binder caller UID via \texttt{Binder.getCallingUid()} or bind access to a trusted signing identity using signature-level permissions or \texttt{PackageManager.checkSignatures()}.

%% file: related_work.tex
\section{Related Work}
\label{related}

\noindent
\textbf{Security and privacy in pre-installed apps.}
There are many efforts that have been devoted to understanding the security and privacy of pre-installed apps in Android.
For example, some studies show that Android firmware remains difficult to audit at scale and continues to contain security inconsistencies across vendors that hinder pre-installed app analysis~\cite{hou2023can,sutter2023firmwaredroid,silver}, some works report vulnerabilities stemming from custom permissions used to assist data sharing along the supply chain (e.g.,~\cite{gamba2020analysis}), and others identify privilege-escalation vulnerabilities (e.g.,~\cite{firmscope}).
While these studies establish pre-installed apps as a particularly high-impact setting for security misconfigurations, they do not evaluate whether enforced access controls are sufficient to avoid unauthorized operations in \texttt{ContentProvider}.

\noindent
\textbf{Security and privacy in Android components.}
The literature has long recognized that exported components in Android are a source of security vulnerabilities (e.g.,~\cite{chin2011analyzing,zhou2024beyond,felt2011permission,lu2012chex}).
In particular, prior work has reported vulnerable access control problems~\cite{chin2011analyzing}, permission re-delegation issues~\cite{felt2011permission}, and component-hijacking vulnerabilities~\cite{lu2012chex}. 
With respect to \texttt{ContentProvider}, the literature has found passive content leaks and pollution issues because of misconfiguration (e.g.,~\cite{jiang2013detecting, lenk2023poster, bagheri2015detection, talegaon2021formal}).
However, these studies primarily focus on exposure and structural validity of access control configurations, rather than on whether the protection level complies with the sensitivity of the exposed operation and data.\looseness=-1

\noindent
\textbf{LLM-assisted approaches in app analysis.}
Recent works have shown that LLMs can perform useful semantic reasoning in adjacent Android security tasks:
SecretLoc~\cite{alecci2025evaluating} used LLMs to detect hardcoded secrets that regex-based tools miss, and PathSentinel~\cite{lee2024static} used LLM-generated exploit code to reduce the cost of validating filesystem vulnerabilities.
These efforts provide methodological evidence that language models can reason about security-relevant semantics.
To our knowledge, prior work does not directly assess whether the protection of a \texttt{ContentProvider} complies with the semantic sensitivity of the operation and data it exposes, particularly in pre-installed apps.
In addition, different from these efforts, \sysname does not use LLMs as a replacement for program analysis; instead, it combines statically extracted program evidence with LLM-based semantic reasoning for analysis.

%% file: conclusion.tex
\section{Conclusion}
\label{sec:conclusion}

Although many efforts have been made to explore vulnerabilities in pre-installed apps, existing analyses largely focus on API-level sensitivity and overlook vulnerabilities whose security impact emerges only at specific callsites.
Our study shows that insufficient access control in \texttt{ContentProvider} components of pre-installed apps represents a systematic and previously underexplored vulnerability class.
Therefore, we propose \sysname, an access control sufficiency analysis, to fill this gap with a context-aware sensitivity assessment that identifies security-sensitive callsites from the operations and data involved, as well as the concept of restriction context to determine whether enforced access controls provide sufficient protection against unauthorized operations.
We evaluated \sysname on $50$ firmware images covering $1{,}750$ distinct pre-installed apps from $10$ top-tier vendors, and uncovered $819$ vulnerable paths across $424$ \texttt{ContentProvider}s spanning all $10$ vendors.
Our findings reveal that these vulnerabilities are not isolated implementation mistakes: most originate from vendor-customized code, and many expose sensitive resources through generic APIs whose security relevance cannot be determined from API identity alone.
As of this writing, the associated vendors have confirmed $278$ of our disclosed reports, demonstrating the practical impact of this overlooked vulnerability class.

%% file: ethics.tex
\section{Ethics Considerations}
\label{sec:ethics}

We carefully consider ethical issues throughout this study and conduct responsible disclosure practices.
First, \sysname analyzes the binaries of pre-installed apps extracted from firmware images entirely offline.
Therefore, the analysis procedures do not interact with vendor backends, production services, or user devices and impose no burden on vendors' operational infrastructure.
The analysis neither interacts with real users nor collects or accesses user-generated data, thereby avoiding risks to user privacy.
During manual verification, we conduct experiments only on our own devices, ensuring that testing does not affect other users or deployed systems.
Finally, we follow responsible disclosure practices and report validated vulnerabilities only to affected vendors through established security channels.
Validation and responsible disclosure of the remaining $327$ paths are ongoing.
We report each path to the affected vendor once it is manually confirmed, and count only confirmed paths in the manual-validation and vendor-confirmation numbers reported here.

%% file: appendix/appendix-prompts.tex
\appendices

\section{Case Studies and Proofs of Concept}
\label{sec:case-study}
\label{app:pocs}

This appendix presents the root cause, vulnerable implementation, exploit path, and corresponding proof of concept (PoC) for three vendor-acknowledged cases.
All tests were conducted only on devices owned by the researchers, as described in \S\ref{sec:ethics}.

\noindent \textit{1)} \textbf{Vendor A: Arbitrary AI instruction execution}

\noindent
\textbf{\textit{Vulnerability and the root cause.}}
This \texttt{ContentProvider} is within Vendor A's pre-installed AI voice assistant, which is shipped on hundreds of millions of devices.
As shown in \autoref{lst:xiaomi-voiceassist}, the externally accessible \texttt{ContentProviderA} relies on two checks in \texttt{call()}, both of which fall into the insecure category in \autoref{tab:android-access-control-modeling}.
First, it compares the caller-supplied \texttt{Bundle} field \texttt{from}, a \emph{Self-Claimed Extra} pattern, with the hard-coded string \texttt{device\_link}, to imitate an external device connection.
Then, it admits the caller when the calling package name, a \emph{Package Name} pattern, appears in an allowlist, or when the caller-supplied \texttt{support\_package} field, a \emph{Caller-Supplied Package} pattern, equals the calling package name.
Both checks are insecure; an unprivileged app can potentially exploit these forgeable checks and finally reach the sensitive callsite, i.e., \texttt{G.dispatch()}.

\noindent
\textbf{\textit{PoC and impact.}}
\autoref{lst:xiaomi-poc} shows our PoC exploitation of the vulnerability.
By setting the caller-controlled \texttt{from} field to \texttt{device\_link} and the \texttt{support\_package} field to the attacker app's own package name, an arbitrary JSON \texttt{instruction} payload from an unprivileged app can reach the highly privileged function \texttt{G.dispatch} and be eventually forwarded to \texttt{VoiceQueryService} without any further authentication.
This enables attackers to invoke all the critical functionalities provided by this AI voice assistant, including remote vehicle control, smart-home actuation, and unauthorized personal messaging.

\vspace{-0.05in}
\begin{figure}[H]
\centering
\begin{tikzpicture}[remember picture]
\node[inner sep=0pt] (codeblock) {%
\begin{minipage}{0.98\columnwidth}
\begin{lstlisting}[language=Java,
caption={Arbitrary AI instruction execution.},
    label={lst:xiaomi-voiceassist},
    escapeinside={(*@}{@*)},
    basicstyle=\fontsize{5.5}{7}\ttfamily]
public final class ContentProviderA extends ContentProvider {
    public Bundle call(String method, String str, Bundle bundle) {
        if (!bundle.getString("from").equals("device_link")) return; (*@\label{line:xiaomi-g1}@*)
        if (!C.isDeviceSupportPackages(getCallingPackage())
                && !bundle.getString("support_package")
                .equals(getCallingPackage())) return; (*@\label{line:xiaomi-g2}@*)
        if (method.equals("sendInstruction"))
            return G.dispatch(bundle, getCallingPackage()); (*@\label{line:xiaomi-dispatch}@*)
    }
}
class G {
    Bundle dispatch(Bundle bundle, String str) {
        String instruction = bundle.getString("instruction"); (*@\label{line:xiaomi-instr}@*)
        Intent intent = new Intent("com.xxx.voiceassist.query");
        intent.setClassName("com.xxx.voiceassist",
            "com.xxx.voiceassistant.VoiceQueryService");
        intent.putExtra("assist_query", instruction);
        VoiceAssistApplication.getContext().startService(intent); (*@\label{line:xiaomi-sink}@*)
    }
}
\end{lstlisting}
\end{minipage}
};
\end{tikzpicture}
 \vspace{-0.25in}
\end{figure}

\vspace{-0.15in}
\begin{figure}[H]
\centering
\begin{tikzpicture}[remember picture]
\node[inner sep=0pt] (codeblock) {%
\begin{minipage}{0.98\columnwidth}
\begin{lstlisting}[language=Java,
    caption={Proof of concept for \autoref{lst:xiaomi-voiceassist}.},
    label={lst:xiaomi-poc},
    escapeinside={(*@}{@*)},
    basicstyle=\fontsize{5.5}{7}\ttfamily]
Bundle extras = new Bundle();
extras.putString("from", "device_link"); (*@\label{line:xiaomi-poc-g1}@*)
extras.putString("support_package", ctx.getPackageName()); (*@\label{line:xiaomi-poc-g2}@*)
extras.putString("instruction", buildInstruction(
    "Phone", "SendMessage", "+15555550100")); (*@\label{line:xiaomi-poc-instr}@*)
Uri uri = Uri.parse("content://com.xxx.voiceassist"
    + ".contentprovider.ContentProviderA");
ctx.getContentResolver().call(
    uri, "sendInstruction", null, extras); (*@\label{line:xiaomi-poc-call}@*)
\end{lstlisting}
\end{minipage}
};
\end{tikzpicture}
\vspace{-0.15in}
\end{figure}

\noindent \textit{2)} \textbf{Vendor B: Plaintext user privacy exposure}

\noindent
\textbf{\textit{Vulnerability and the root cause.}}
This pre-installed app ships with Vendor B's firmware, which runs on hundreds of millions of devices.
As presented in \autoref{lst:oppo-dcs-query}, it contains a \texttt{ContentProvider}, \texttt{ContentProviderB}, that is declared as externally accessible without any component-level permission.
However, its \texttt{query} interface only performs a fragile package-name check on a caller-controlled \texttt{URI} parameter with a hard-coded string.

\noindent
\textbf{\textit{PoC and impact.}}
\autoref{lst:oppo-poc} shows our PoC attack on the discovered vulnerability.
By appending the forged \texttt{URI} query parameter \texttt{packageName=com.xxx.yyy}, an unprivileged app can bypass the access control in \texttt{ContentProviderB.query()} and reach \texttt{h1.k1}, which directly queries the plaintext \texttt{common\_info} table and returns the resulting \texttt{Cursor} without further permission enforcement.
Since this data collection service centrally stores raw telemetry from privileged components,
in our PoC, the attacker can enumerate plaintext entries containing full voice-assistant queries, precise GPS coordinates and photo paths, SMS URLs, and device or account-related identifiers, resulting in severe privacy breaches.\looseness=-1

\vspace{-0.05in}
\begin{figure}[H]
\centering
\begin{tikzpicture}[remember picture]
\node[inner sep=0pt] (codeblock) {%
\begin{minipage}{0.98\columnwidth}
\begin{lstlisting}[language=Java,
    caption={Plaintext user privacy exposure.},
    label={lst:oppo-dcs-query},
    escapeinside={(*@}{@*)},
    basicstyle=\fontsize{5.5}{7}\ttfamily]
public class ContentProviderB extends ContentProvider {
    public Cursor query(Uri uri, String[] projection,
                        String sel, String[] selArgs, ...) {
        String pkg = uri.getQueryParameter("packageName"); (*@\label{line:oppo-pkg}@*)
        if (TextUtils.equals("com.xxx.yyy", pkg)) { (*@\label{line:oppo-g1}@*)
            return h1.k1(projection, sel, selArgs, sortOrder); (*@\label{line:oppo-dispatch}@*)
        }
        return null;
    }
}
class h1 {
    public static Cursor k1(String[] projection, String sel,
                             String[] selArgs, ...) {
        return db.query(true, "common_info", (*@\label{line:oppo-sink}@*)
                        projection, sel, selArgs,
                        null, null, sortOrder, null);
    }
}
\end{lstlisting}
\end{minipage}
};
\end{tikzpicture}
\vspace{-0.25in}
\end{figure}

\vspace{-0.15in}
\begin{figure}[H]
\centering
\begin{tikzpicture}[remember picture]
\node[inner sep=0pt] (codeblock) {%
\begin{minipage}{0.98\columnwidth}
\begin{lstlisting}[language=Java,
    caption={Proof of concept for \autoref{lst:oppo-dcs-query}.},
    label={lst:oppo-poc},
    escapeinside={(*@}{@*)},
    basicstyle=\fontsize{5.5}{7}\ttfamily]
Uri uri = Uri.parse("content://com.xxx.statistics.provider"
    + "/query_event?packageName=com.xxx.yyy"); (*@\label{line:oppo-poc-uri}@*)

Cursor c = ctx.getContentResolver().query(uri, null, null, null, null);(*@\label{line:oppo-poc-query}@*)

while (c != null && c.moveToNext()) {
    Log.d(TAG, c.getString(c.getColumnIndexOrThrow("common_body"))); (*@\label{line:oppo-poc-read}@*)
}
\end{lstlisting}
\end{minipage}
};
\end{tikzpicture}
\vspace{-0.15in}
\end{figure}

\noindent \textit{3)} \textbf{Vendor C: Privacy agreement acceptance enforcement}

\noindent
\textbf{\textit{Vulnerability and the root cause.}}
This pre-installed app is Vendor C's game assistant, which is shipped on hundreds of millions of devices, provides game-related services, and maintains a privacy-agreement state for its agreement-gated functionality.
Its \texttt{ContentProviderC} is part of the agreement AIDL module and acts as an IPC bridge that returns an \texttt{IAgreeService} Binder to callers.
However, this provider is declared as externally accessible, and its declared \texttt{readPermission} and \texttt{writePermission} are both defined with a self-granted protection level \texttt{normal}.
As a result, any third-party app can obtain these permissions by declaring them in its manifest without user approval or privileged signing.
As shown in \autoref{lst:honor-agreeprovider}, the remaining access control in \texttt{call()} only checks whether \texttt{method} equals \texttt{getService} and \texttt{arg} equals the hardcoded magic string \texttt{xxxxxxxx}.
Both values are static constants embedded in the APK and can be recovered through static analysis.
Once this forgeable check is satisfied, \texttt{ContentProviderC} returns the raw \texttt{AgreeServiceStub} Binder through \texttt{BundleCompat.putBinder()}.

\noindent
\textbf{\textit{PoC and impact.}}
\autoref{lst:honor-poc} shows our PoC exploitation.
By invoking \texttt{ContentResolver.call()} with \texttt{method=getService} and a payload with a hardcoded string \texttt{arg}, an unprivileged app that declares the \texttt{normal}-level provider permission, which is self-granted, can obtain the exposed \texttt{AgreeServiceStub} Binder.
The attacker can then invoke the Binder transaction corresponding to \texttt{signUdid()}, which updates the local agreement state and writes \texttt{game\_agreement\_sign\_state\_<area> = 1} into \texttt{Settings.Global}.
This primitive effectively creates a backdoor-like consent bypass in the privacy-agreement workflow, allowing a third-party app to force this agreement state without any user interaction.
Therefore, the vulnerability can silently convert a user-controlled consent decision into an attacker-triggered state change, potentially enabling the collection and processing of user data without the user's knowledge.\looseness=-1

\vspace{-0.05in}
\begin{figure}[H]
\centering
\begin{tikzpicture}[remember picture]
\node[inner sep=0pt] (codeblock) {%
\begin{minipage}{0.98\columnwidth}
\begin{lstlisting}[language=Java,
    caption={Privacy agreement acceptance enforcement.},
    label={lst:honor-agreeprovider},
    escapeinside={(*@}{@*)},
    basicstyle=\fontsize{5.5}{7}\ttfamily]
public class ContentProviderC extends ContentProvider {
    public Bundle call(String method, String arg, Bundle bundle) {
        if (!"getService".equals(method) (*@\label{line:honor-g1}@*)
                || !"xxxxxxxx".equals(arg)) return null; (*@\label{line:honor-g2}@*)
        Bundle result = new Bundle();
        BundleCompat.putBinder(result, "binder", this.f3277a); (*@\label{line:honor-sink}@*)
        return result;
    }
}
class AgreeServiceStub extends IAgreeService.Stub {
    public void signUdid() throws RemoteException { (*@\label{line:honor-signudid}@*)
        Settings.Global.putInt(context.getContentResolver(), (*@\label{line:honor-global}@*)
            "game_agreement_sign_state_".concat(area), 1);
    }
}
\end{lstlisting}
\end{minipage}
};
\end{tikzpicture}
\vspace{-0.25in}
\end{figure}

\vspace{-0.15in}
\begin{figure}[H]
\centering
\begin{minipage}{0.98\columnwidth}
\begin{lstlisting}[language=Java,
    caption={Proof of concept for \autoref{lst:honor-agreeprovider}.},
    label={lst:honor-poc},
    escapeinside={(*@}{@*)},
    basicstyle=\fontsize{5.5}{7}\ttfamily]
Uri uri = Uri.parse(
    "content://com.xxx.gameassistant.provider");
Bundle result = ctx.getContentResolver().call(
    uri, "getService", "xxxxxxxx", null); (*@\label{line:honor-poc-auth}@*)
IBinder binder = result.getBinder("binder"); (*@\label{line:honor-poc-binder}@*)
Parcel data = Parcel.obtain(), reply = Parcel.obtain();
data.writeInterfaceToken(
    "com.xxx.gameassistant.IAgreeService");
binder.transact(2, data, reply, 0); (*@\label{line:honor-poc-sink}@*)
reply.readException();
\end{lstlisting}
\end{minipage}
\vspace{-0.15in}
\end{figure}

\clearpage
\onecolumn
\section{Detailed Prompt Specifications}
\label{app:prompts}

This appendix provides the detailed prompts used by the three reasoning stages described in \S\ref{sec:design}.

\begin{center}
\begin{minipage}[t]{0.485\textwidth}
\vspace{0pt}
\centering
\input{prompts/prompt_module1}
\captionsetup{font=footnotesize,skip=3pt}
\captionof{figure}{Stage 1 prompt for callsite-sensitivity classification.}
\label{fig:prompt-module1}
\end{minipage}
\hfill
\begin{minipage}[t]{0.485\textwidth}
\vspace{0pt}
\centering
\input{prompts/prompt_module2}
\captionsetup{font=footnotesize,skip=3pt}
\captionof{figure}{Stage 2 prompt for branch-check semantic classification.}
\label{fig:prompt-module2}
\end{minipage}
\end{center}

\vspace{0.14in}

\begin{center}
\input{prompts/prompt_module3}
\captionsetup{font=footnotesize,skip=3pt}
\captionof{figure}{Stage 3 prompt for protection coverage and sufficiency assessment.}
\label{fig:prompt-module3}
\end{center}

\clearpage

%% file: prompts/prompt_module1.tex
\definecolor{hlFocus}{RGB}{245,232,210}
\definecolor{hlTok}{RGB}{220,228,238}

\begin{tcolorbox}[
enhanced,
sharp corners,
boxrule=0.3pt,
colback=white,
colframe=black!60,
left=3pt,
right=3pt,
top=2pt,
bottom=2pt,
height=0.47\textheight,
fontupper=\scriptsize,
fontlower=\scriptsize,
title={\small\faFile\ \textbf{Stage 1 Prompt: Callsite Sensitivity}\hfill \scriptsize \S\ref{sec:design:p1}},
coltitle=black,
colbacktitle=black!6,
segmentation style={solid, black!60, line width=0.3pt}
]

Decide the \tightbox{hlFocus}{sensitivity} of the following callsite to an Android system API.\\
You are given two kinds of analysis evidence:
\tightbox{hlTok}{API signature} and \tightbox{hlTok}{argument backward slices}.
The backward slice of each argument explains where the argument value comes from.\\[2pt]

\textbf{Operation/data taxonomy}:\\
Use the operation categories and data categories defined in \autoref{tab:sensitive_matrix}.\\
The operation category must be one of $O1$--$O3$.
The data category must be one of $P1$--$P4$ or $U1$--$U3$.\\[2pt]

\textbf{Steps}:\\
\begin{tabularx}{\linewidth}{@{}>{\bfseries}p{3.7em}@{\hspace{2pt}}>{\raggedright\arraybackslash}X@{}}
Step 1: & Identify the implemented operation and affected data using the operation/data taxonomy.\\
Step 2: & Decide whether the callsite is \texttt{sensitive} or \texttt{nonsensitive}, and give the reasoning.\\
Step 3: & Return the JSON object specified below, grounded in the slices.
\end{tabularx}

\tcblower

\textbf{Output Contract (strict JSON)}:\\[-1pt]
{\ttfamily
\{\\
\hspace*{1em}"sensitivity": "\textless sensitive|nonsensitive\textgreater",\\
\hspace*{1em}"reasoning": "\textless ...\textgreater",\\
\hspace*{1em}"sensitive\_exec": \{\\
\hspace*{2em}"operation": "\textless concrete operation or null\textgreater",\\
\hspace*{2em}"operation\_category": "\textless $O1$|$O2$|$O3$|null\textgreater",\\
\hspace*{2em}"data": "\textless affected data/resource or null\textgreater",\\
\hspace*{2em}"data\_category": "\textless $P1$|$P2$|$P3$|$P4$|$U1$|$U2$|$U3$|null\textgreater"\\
\hspace*{1em}\},\\
\hspace*{1em}"evidence": [\\
\hspace*{2em}\{\\
\hspace*{3em}"arg\_index": "\textless integer\textgreater",\\
\hspace*{3em}"arg\_slice": "\textless cited backward slice\textgreater"\\
\hspace*{2em}\}\\
\hspace*{1em}]\\
\}\par}

\end{tcolorbox}

%% file: prompts/prompt_module2.tex
\definecolor{hlFocus}{RGB}{245,232,210}
\definecolor{hlTok}{RGB}{220,228,238}
\begin{tcolorbox}[
enhanced,
sharp corners,
boxrule=0.3pt,
colback=white,
colframe=black!60,
left=3pt,
right=3pt,
top=2pt,
bottom=2pt,
height=0.47\textheight,
fontupper=\scriptsize,
fontlower=\scriptsize,
title={\small\faFile\ \textbf{Stage 2 Prompt: Check Semantics}\hfill \scriptsize \S\ref{sec:design:p2}},
coltitle=black,
colbacktitle=black!6,
segmentation style={solid, black!60, line width=0.3pt}
]

Your task is to classify the \tightbox{hlFocus}{security purpose} of this branch condition.
You should infer what the branch condition is checking and how it performs the check.\\[1pt]

You are given \tightbox{hlTok}{branch condition backward slices} of one branch condition expression extracted from a sensitive path by \S\ref{sec:design:p1}.
The condition slices consist of:
(1) the original branch condition expression;
(2) the backward slice of the left-hand side; and
(3) the backward slice of the right-hand side.\\[2pt]
Preserve the exact comparison semantics; do not collapse prefix, exact, and case-insensitive equality checks.\\[2pt]

\textbf{Steps}:\\
\begin{tabularx}{\linewidth}{@{}>{\bfseries}p{3.7em}@{\hspace{2pt}}>{\raggedright\arraybackslash}X@{}}
Step 1: & Inspect the condition expression and its left- and right-hand-side slices.\\
Step 2: & Identify what both values represent and where each originates.\\
Step 3: & Record the exact comparison API or operator, distinguishing, for example, \texttt{startsWith}, \texttt{equals}, and \texttt{equalsIgnoreCase}.\\
Step 4: & Classify the purpose as authentication, authorization, or other, using only the supplied slice evidence.\\
Step 5: & Return strict JSON grounded in the slices.
\end{tabularx}\\[-3pt]

\tcblower

\textbf{Output Contract (strict JSON)}:\\[-1pt]
{\ttfamily
\{\\
\hspace*{1em}"purpose": "\textless authentication|authorization|other\textgreater",\\
\hspace*{1em}"reasoning": "\textless ...\textgreater",\\
\hspace*{1em}"mechanism": \{\\
\hspace*{2em}"property": "\textless values or properties being compared\textgreater",\\
\hspace*{2em}"operation": "\textless exact comparison API or operator\textgreater",\\
\hspace*{2em}"data": ["\textless source of each compared value\textgreater"]\\
\hspace*{1em}\},\\
\hspace*{1em}"evidence": \{\\
\hspace*{2em}"expr": "\textless original branch condition expression\textgreater",\\
\hspace*{2em}"lhs\_slice": "\textless left-hand-side slice\textgreater",\\
\hspace*{2em}"rhs\_slice": "\textless right-hand-side slice\textgreater"\\
\hspace*{1em}\}\\
\}\par}
\end{tcolorbox}

%% file: prompts/prompt_module3.tex
\definecolor{hlFocus}{RGB}{245,232,210}
\definecolor{hlTok}{RGB}{220,228,238}

\begin{tcolorbox}[
enhanced,
sharp corners,
boxrule=0.3pt,
colback=white,
colframe=black!60,
left=3pt,
right=3pt,
top=2pt,
bottom=2pt,
fontupper=\scriptsize,
fontlower=\scriptsize,
title={\small\faFile\ \textbf{Stage 3 Prompt: Protection Sufficiency}\hfill \scriptsize \S\ref{sec:design:p3}},
coltitle=black,
colbacktitle=black!6,
segmentation style={solid, black!60, line width=0.3pt},
sidebyside,
sidebyside align=top,
lefthand width=0.54\linewidth,
sidebyside gap=8pt,
]

Assess whether the sensitive execution is properly protected by the manifest authorization facts and the authentication or authorization checks on this path.\\
Produce the following two verdicts: \tightbox{hlFocus}{coverage}, whether any manifest authorization or path check covers the sensitive execution;
\tightbox{hlFocus}{sufficiency}, whether at least one covering protection is strong enough to protect that sensitive execution.\\[2pt]

You are given three kinds of input facts: (1) \tightbox{hlTok}{callsite-side output}: \texttt{sensitive\_exec}, \texttt{reasoning}, callsite evidence, and \texttt{call\_chain};
(2) \tightbox{hlTok}{manifest authorization facts}: permissions or access restrictions declared for the component, operation axis, or path;
(3) \tightbox{hlTok}{path check outputs}: each dynamic check's \texttt{purpose}, \texttt{reasoning}, \texttt{mechanism}, and branch-condition evidence.
All path check outputs are authentication or authorization checks on this path.\\[2pt]

\textbf{Steps}:\\
\begin{tabularx}{\linewidth}{@{}>{\bfseries}p{3.7em}@{\hspace{2pt}}>{\raggedright\arraybackslash}X@{}}
Step 1: & Identify the sensitive execution from the callsite-side output.\\
Step 2: & For each protection, set \texttt{covers=true} only if it constrains the same operation and data as the sensitive execution.\\
Step 3: & For each covering protection, set \texttt{sufficient=true} only if its authority cannot be controlled or forged by the attacker; otherwise set it to \texttt{false}. For a non-covering protection, set \texttt{sufficient=null}.\\
Step 4: & Set global \texttt{coverage=covered} iff any protection covers. Set global \texttt{sufficiency=sufficient} iff any covering protection is sufficient, \texttt{insufficient} if protections cover but none is sufficient, and \texttt{not\_applied} if none covers.\\
Step 5: & Return strict JSON grounded in the callsite-side, manifest authorization facts, and path-check evidence.
\end{tabularx}\\[-1pt]

\tcblower

\textbf{Output Contract (strict JSON)}:\\[-1pt]
{\ttfamily
\{\\
\hspace*{1em}"coverage": "\textless covered|uncovered\textgreater",\\
\hspace*{1em}"sufficiency":\\
\hspace*{2em}"\textless sufficient|insufficient|not\_applied\textgreater",\\
\hspace*{1em}"reasoning": "\textless ...\textgreater",\\
\hspace*{1em}"sensitive\_exec": "\textless ...\textgreater",\\
\hspace*{1em}"call\_chain": [\\
\hspace*{2em}\{\\
\hspace*{3em}"method\_sign": "\textless method signature\textgreater",\\
\hspace*{3em}"call\_index": "\textless integer or null\textgreater"\\
\hspace*{2em}\}\\
\hspace*{1em}],\\
\hspace*{1em}"protections": [\\
\hspace*{2em}\{\\
\hspace*{3em}"covers": "\textless true|false\textgreater",\\
\hspace*{3em}"sufficient": "\textless true|false|null\textgreater",\\
\hspace*{3em}"reasoning": "\textless ...\textgreater",\\
\hspace*{3em}"source": "\textless manifest|path\_check\textgreater",\\
\hspace*{3em}"purpose":\\
\hspace*{4em}"\textless authorization|authentication\textgreater",\\
\hspace*{3em}"mechanism": "\textless protection summary\textgreater"\\
\hspace*{2em}\}\\
\hspace*{1em}]\\
\}\par}

\end{tcolorbox}

%% file: main.bbl
\begin{thebibliography}{10}
\providecommand{\url}[1]{#1}
\csname url@samestyle\endcsname
\providecommand{\newblock}{\relax}
\providecommand{\bibinfo}[2]{#2}
\providecommand{\BIBentrySTDinterwordspacing}{\spaceskip=0pt\relax}
\providecommand{\BIBentryALTinterwordstretchfactor}{4}
\providecommand{\BIBentryALTinterwordspacing}{\spaceskip=\fontdimen2\font plus
\BIBentryALTinterwordstretchfactor\fontdimen3\font minus
  \fontdimen4\font\relax}
\providecommand{\BIBforeignlanguage}[2]{{%
\expandafter\ifx\csname l@#1\endcsname\relax
\typeout{** WARNING: IEEEtran.bst: No hyphenation pattern has been}%
\typeout{** loaded for the language `#1'. Using the pattern for}%
\typeout{** the default language instead.}%
\else
\language=\csname l@#1\endcsname
\fi
#2}}
\providecommand{\BIBdecl}{\relax}
\BIBdecl

\bibitem{wu2013impact}
L.~Wu, M.~Grace, Y.~Zhou, C.~Wu, and X.~Jiang, ``The impact of vendor
  customizations on {Android} security,'' in \emph{Proceedings of ACM CCS},
  2013.

\bibitem{harvesting16}
Y.~Aafer, X.~Zhang, and W.~Du, ``Harvesting inconsistent security
  configurations in custom {Android} {ROMs} via differential analysis,'' in
  \emph{Proceedings of USENIX Security}, 2016.

\bibitem{gamba2020analysis}
J.~Gamba, M.~Rashed, A.~Razaghpanah, J.~Tapiador, and N.~Vallina-Rodriguez,
  ``An analysis of pre-installed {Android} software,'' in \emph{Proceedings of
  IEEE S\&P}, 2020.

\bibitem{firmscope}
M.~Elsabagh, R.~Johnson, A.~Stavrou, C.~Zuo, Q.~Zhao, and Z.~Lin,
  ``{FirmScope}: Automatic uncovering of privilege-escalation vulnerabilities
  in pre-installed apps in {Android} firmware,'' in \emph{Proceedings of USENIX
  Security}, 2020.

\bibitem{patch22}
Q.~Hou, W.~Diao, Y.~Wang, X.~Liu, S.~Liu, L.~Ying, S.~Guo, Y.~Li, M.~Nie, and
  H.~Duan, ``Large-scale security measurements on the {Android} firmware
  ecosystem,'' in \emph{Proceedings of IEEE/ACM ICSE}, 2022.

\bibitem{news}
\BIBentryALTinterwordspacing
{Heise}. (2025) Pre-installed apps with security vulnerabilities on low-cost
  smartphones. Accessed: May 7, 2026. [Online]. Available:
  \url{https://www.heise.de/en/news/Pre-installed-apps-with-security-vulnerabilities-on-low-cost-smartphones-10423996.html}
\BIBentrySTDinterwordspacing

\bibitem{zhao2020automatic}
Q.~Zhao, C.~Zuo, B.~Dolan-Gavitt, G.~Pellegrino, and Z.~Lin, ``Automatic
  uncovering of hidden behaviors from input validation in mobile apps,'' in
  \emph{Proceedings of IEEE S\&P}, 2020.

\bibitem{yang2013appintent}
Z.~Yang, M.~Yang, Y.~Zhang, G.~Gu, P.~Ning, and X.~S. Wang, ``{AppIntent}:
  Analyzing sensitive data transmission in {Android} for privacy leakage
  detection,'' in \emph{Proceedings of ACM CCS}, 2013.

\bibitem{arzt2015using}
S.~Arzt, S.~Rasthofer, R.~Hahn, and E.~Bodden, ``Using targeted symbolic
  execution for reducing false-positives in dataflow analysis,'' in
  \emph{Proceedings of ACM SOAP}, 2015.

\bibitem{cadar2013symbolic}
C.~Cadar and K.~Sen, ``Symbolic execution for software testing: three decades
  later,'' \emph{Communications of the ACM}, vol.~56, no.~2, pp. 82--90, 2013.

\bibitem{android-contentprovider-doc}
\BIBentryALTinterwordspacing
Android. (2026) {Android} {ContentProvider} {API} reference. Accessed: May 7,
  2026. [Online]. Available:
  \url{https://developer.android.com/reference/android/content/ContentProvider}
\BIBentrySTDinterwordspacing

\bibitem{aospprovider}
\BIBentryALTinterwordspacing
------. (2026) {AOSP}: {ContentProvider} source code. Accessed: May 7, 2026.
  [Online]. Available:
  \url{https://cs.android.com/android/platform/superproject/main/+/main:frameworks/base/core/java/android/content/ContentProvider.java}
\BIBentrySTDinterwordspacing

\bibitem{android-system-permission-doc}
\BIBentryALTinterwordspacing
{Android Developers}. (2026) {Android} permissions overview. Accessed: May 7,
  2026. [Online]. Available:
  \url{https://developer.android.com/guide/topics/permissions/overview}
\BIBentrySTDinterwordspacing

\bibitem{provider}
\BIBentryALTinterwordspacing
------. (2026) {\texttt{<provider>}}: {Android} manifest element reference.
  Accessed: May 7, 2026. [Online]. Available:
  \url{https://developer.android.com/guide/topics/manifest/provider-element}
\BIBentrySTDinterwordspacing

\bibitem{android-context-api-doc}
\BIBentryALTinterwordspacing
------. (2026) {android.content.Context API Reference}. Accessed: May 7, 2026.
  [Online]. Available:
  \url{https://developer.android.com/reference/android/content/Context}
\BIBentrySTDinterwordspacing

\bibitem{flowdroid}
S.~Arzt, S.~Rasthofer, C.~Fritz, E.~Bodden, A.~Bartel, J.~Klein, Y.~Le~Traon,
  D.~Octeau, and P.~McDaniel, ``{FlowDroid}: Precise context, flow, field,
  object-sensitive and lifecycle-aware taint analysis for {Android} apps,'' in
  \emph{Proceedings of ACM PLDI}, 2014.

\bibitem{wei2018amandroid}
F.~Wei, S.~Roy, X.~Ou, and Robby, ``{Amandroid}: A precise and general
  inter-component data flow analysis framework for security vetting of
  {Android} apps,'' \emph{{ACM} Transactions on Privacy and Security}, vol.~21,
  no.~3, pp. 1--32, 2018.

\bibitem{zuo2019leakscope}
C.~Zuo, Z.~Lin, and Y.~Zhang, ``Why does your data leak? {Uncovering} the data
  leakage in cloud from mobile apps,'' in \emph{Proceedings of IEEE S\&P},
  2019.

\bibitem{android-severity}
\BIBentryALTinterwordspacing
Android. (2026) {Google} severity criteria. Accessed: May 7, 2026. [Online].
  Available:
  \url{https://source.android.com/docs/security/overview/updates-resources}
\BIBentrySTDinterwordspacing

\bibitem{samsung-severity}
\BIBentryALTinterwordspacing
{Samsung}. (2026) {Samsung} severity criteria. Accessed: May 7, 2026. [Online].
  Available:
  \url{https://security.samsungmobile.com/securityReportingRiskClassification.smsb}
\BIBentrySTDinterwordspacing

\bibitem{xiaomi-severity}
\BIBentryALTinterwordspacing
{Xiaomi}. (2026) {Xiaomi} severity criteria. Accessed: May 7, 2026. [Online].
  Available: \url{https://sec.xiaomi.com/#/notice/detail/228}
\BIBentrySTDinterwordspacing

\bibitem{ma2023lms}
W.~Ma, Z.~Lin, S.~Liu, Q.~Hu, Y.~Liu, W.~Wang, C.~Zhang, L.~Nie, L.~Li, Y.~Liu,
  and L.~Jiang, ``Exploring code analysis: Zero-shot insights on syntax and
  semantics with {LLMs},'' \emph{ACM Transactions on Software Engineering and
  Methodology}, 2025, \url{https://doi.org/10.1145/3818607}.

\bibitem{nam2024using}
D.~Nam, A.~Macvean, V.~Hellendoorn, B.~Vasilescu, and B.~Myers, ``Using an
  {LLM} to help with code understanding,'' in \emph{Proceedings of IEEE/ACM
  ICSE}, 2024.

\bibitem{wang2024llmdfa}
C.~Wang, W.~Zhang, Z.~Su, X.~Xu, X.~Xie, and X.~Zhang, ``{LLMDFA}: Analyzing
  dataflow in code with large language models,'' in \emph{Proceedings of
  NeurIPS}, 2024.

\bibitem{wang2025can}
W.~Wang, K.~Liu, Z.~Sun, A.~R. Chen, G.~Li, G.~Huang, and L.~Ma, ``Can large
  language models solve path constraints in symbolic execution?'' \emph{arXiv
  preprint arXiv:2511.18288}, 2025.

\bibitem{cwe862}
\BIBentryALTinterwordspacing
{MITRE}. (2026) {CWE-862}: Missing authorization. Accessed: May 7, 2026.
  [Online]. Available: \url{https://cwe.mitre.org/data/definitions/862.html}
\BIBentrySTDinterwordspacing

\bibitem{cwe863}
\BIBentryALTinterwordspacing
------. (2026) {CWE-863}: Incorrect authorization. Accessed: May 7, 2026.
  [Online]. Available: \url{https://cwe.mitre.org/data/definitions/863.html}
\BIBentrySTDinterwordspacing

\bibitem{soot}
R.~Vall{\'e}e-Rai, P.~Co, E.~Gagnon, L.~Hendren, P.~Lam, and V.~Sundaresan,
  ``Soot: A {Java} bytecode optimization framework,'' in \emph{Proceedings of
  CASCON}, 1999, \url{https://github.com/soot-oss/soot}.

\bibitem{deepseek}
\BIBentryALTinterwordspacing
{DeepSeek}. (2026) {DeepSeek} {API} documentation. Accessed: May 7, 2026.
  [Online]. Available: \url{https://platform.deepseek.com/}
\BIBentrySTDinterwordspacing

\bibitem{langchain}
\BIBentryALTinterwordspacing
{LangChain}. (2026) {LangChain} {API} documentation. Accessed: May 7, 2026.
  [Online]. Available: \url{https://www.langchain.com/}
\BIBentrySTDinterwordspacing

\bibitem{topoem}
\BIBentryALTinterwordspacing
{AppBrain}. (2025) Top {Android} phone manufacturers. Accessed: May 7, 2026.
  [Online]. Available: \url{https://www.appbrain.com/stats/top-manufacturers}
\BIBentrySTDinterwordspacing

\bibitem{androiddumps}
\BIBentryALTinterwordspacing
{Android Dumps}. (2026) {Android Dumps}. Accessed: May 7, 2026. [Online].
  Available: \url{https://dumps.tadiphone.dev/dumps}
\BIBentrySTDinterwordspacing

\bibitem{firmwaredrive}
\BIBentryALTinterwordspacing
{FirmwareDrive}. (2026) Database of stock firmware, drivers, and tools.
  Accessed: May 7, 2026. [Online]. Available: \url{https://firmwaredrive.com/}
\BIBentrySTDinterwordspacing

\bibitem{tygr}
C.~Zhu, Z.~Li, A.~Xue, A.~P. Bajaj, W.~Gibbs, Y.~Liu, R.~Alur, T.~Bao, H.~Dai,
  A.~Doup\'{e}, M.~Naik, Y.~Shoshitaishvili, R.~Wang, and A.~Machiry, ``{TYGR}:
  Type inference on stripped binaries using graph neural networks,'' in
  \emph{Proceedings of USENIX Security}, 2024.

\bibitem{carlini2026zerodays}
N.~Carlini, K.~Lucas, E.~Ben~Asher, N.~Cheng, H.~Lakhani, D.~Forsythe, and
  K.~Guru, ``Evaluating and mitigating the growing risk of {LLM}-discovered
  0-days,'' Anthropic Frontier Red Team, Feb. 2026,
  \url{https://red.anthropic.com/2026/zero-days/}.

\bibitem{hou2023can}
Q.~Hou, W.~Diao, Y.~Wang, C.~Mao, L.~Ying, S.~Liu, X.~Liu, Y.~Li, S.~Guo,
  M.~Nie, and H.~Duan, ``Can we trust the phone vendors? {Comprehensive}
  security measurements on the {Android} firmware ecosystem,'' \emph{{IEEE}
  TSE}, vol.~49, no.~7, pp. 3901--3921, 2023.

\bibitem{sutter2023firmwaredroid}
T.~Sutter and B.~Tellenbach, ``{FirmwareDroid}: Towards automated static
  analysis of pre-installed {Android} apps,'' in \emph{Proceedings of IEEE/ACM
  MOBILESoft}, 2023.

\bibitem{silver}
C.~Wang, Y.~Zhao, J.~Deng, and H.~Wang, ``Born with a silver spoon: On the
  (in)security of native granted app privileges in custom {Android} {ROMs},''
  in \emph{Proceedings of {IEEE} S\&P}, 2025.

\bibitem{chin2011analyzing}
E.~Chin, A.~P. Felt, K.~Greenwood, and D.~Wagner, ``Analyzing inter-application
  communication in {Android},'' in \emph{Proceedings of ACM MobiSys}, 2011.

\bibitem{zhou2024beyond}
H.~Zhou, S.~Wu, C.~Qian, X.~Luo, H.~Cai, and C.~Zhang, ``Beyond the surface:
  Uncovering the unprotected components of {Android} against overlay attack,''
  in \emph{Proceedings of ISOC NDSS}, 2024.

\bibitem{felt2011permission}
A.~P. Felt, H.~J. Wang, A.~Moshchuk, S.~Hanna, and E.~Chin, ``Permission
  re-delegation: Attacks and defenses,'' in \emph{Proceedings of USENIX
  Security}, 2011.

\bibitem{lu2012chex}
L.~Lu, Z.~Li, Z.~Wu, W.~Lee, and G.~Jiang, ``{CHEX}: Statically vetting
  {Android} apps for component hijacking vulnerabilities,'' in
  \emph{Proceedings of ACM CCS}, 2012.

\bibitem{jiang2013detecting}
Y.~Zhou and X.~Jiang, ``Detecting passive content leaks and pollution in
  {Android} applications,'' in \emph{Proceedings of ISOC NDSS}, 2013.

\bibitem{lenk2023poster}
C.~Lenk and J.~Kinder, ``Poster: Privacy risks from misconfigured {Android}
  content providers,'' in \emph{Proceedings of ACM CCS}, 2023.

\bibitem{bagheri2015detection}
H.~Bagheri, E.~Kang, S.~Malek, and D.~Jackson, ``Detection of design flaws in
  the {Android} permission protocol through bounded verification,'' in
  \emph{Proceedings of FM}, 2015.

\bibitem{talegaon2021formal}
S.~Talegaon and R.~Krishnan, ``A formal specification of access control in
  {Android} with {URI} permissions,'' \emph{Information Systems Frontiers},
  vol.~23, no.~4, pp. 849--866, 2021.

\bibitem{alecci2025evaluating}
M.~Alecci, J.~Samhi, T.~F. Bissyand{\'e}, and J.~Klein, ``Evaluating large
  language models in detecting secrets in {Android} apps,'' \emph{arXiv
  preprint arXiv:2510.18601}, 2025.

\bibitem{lee2024static}
Y.-T. Lee, H.~Vijayakumar, Z.~Qian, and T.~Jaeger, ``Static detection of
  filesystem vulnerabilities in {Android} systems,'' \emph{arXiv preprint
  arXiv:2407.11279}, 2024.

\end{thebibliography}
